\documentclass[sigconf]{acmart}

\makeatletter                   
\def\mdseries@tt{m}             
\makeatother                    
\usepackage[draft=true]{minted} 
\usepackage{color}
\usepackage{hyperref}           
\hypersetup{
    colorlinks=true,
    linkcolor=blue,
    filecolor=red,      
    urlcolor=magenta,
    breaklinks=true,            
}
\usepackage{breakurl}           

\usepackage[utf8]{inputenc}

\usepackage{amsmath}
\usepackage{amssymb}
\usepackage{array}
\usepackage{arydshln}
\usepackage{booktabs}
\usepackage{collcell}
\usepackage{comment}
\usepackage{filecontents}
\usepackage{float}
\usepackage{ifthenx}
\usepackage{lipsum}
\usepackage{makecell}
\usepackage{multirow}
\usepackage{nicefrac}
\usepackage{paralist}
\usepackage{flushend}
\usepackage{upgreek}
\usepackage{siunitx}
\usepackage{todonotes}
\usepackage{threeparttable}
\PassOptionsToPackage{hyphens}{url}\usepackage{url}
\usepackage{xspace}
\usepackage{placeins}

\def\subheading#1{\medskip\noindent{\boldmath\textbf{#1}}~\ignorespaces}

\usepackage{tikz}
\usepackage{pgfplots}
\usetikzlibrary{pgfplots.groupplots}
\usetikzlibrary{arrows}
\usetikzlibrary{patterns}
\usetikzlibrary{positioning}
\usetikzlibrary{decorations.pathreplacing}
\usetikzlibrary{shapes.arrows}
\usetikzlibrary{pgfplots.groupplots}
\pgfplotsset{compat=1.8}

\usepackage{caption}
\usepackage{subcaption}
\usepackage{letltxmacro}
\reversemarginpar
\usepackage{todonotes}
\usepackage{marginnote} 
\let\marginpar\marginnote
\LetLtxMacro{\oldtodo}{\todo}
\renewcommand{\todo}[2][]{\tikzexternaldisable\oldtodo[fancyline,size=\footnotesize,#1]{#2}\tikzexternalenable}
\renewcommand{\todo}[1]{\tikzexternaldisable\oldtodo[fancyline,size=\footnotesize]{#1}\tikzexternalenable}

\usepackage{algorithmic}
\usepackage[ruled]{algorithm2e}
\usepackage{listings}
\newfloat{listing}{tbhp}{lst}
\floatname{listing}{Listing}
\lstdefinelanguage
   [x64]{Assembler}     
   [x86masm]{Assembler} 
   {morekeywords={CDQE,CQO,CMPSQ,CMPXCHG16B,JRCXZ,LODSQ,MOVSXD, %
                  POPFQ,PUSHFQ,SCASQ,STOSQ,IRETQ,RDTSCP,SWAPGS, %
                  rax,rdx,rcx,rbx,rsi,rdi,rsp,rbp, %
                  r8,r8d,r8w,r8b,r9,r9d,r9w,r9b}} 

\newcommand{\SIx}[1]{\num{#1}\relax}

\newcommand{\etal}{et~al.\ } 
\newcommand{\ie}{\textit{i.e.},\ } 
\newcommand{\eg}{e.g.,\ } 
\newcommand{\cf}{cf.\ } 

\newcommand{\FlushReload}{Flush+\allowbreak Reload\xspace}

\newcommand{\FlushFlush}{Flush+\allowbreak Flush\xspace}

\newcommand{\DropIt}{DropIt\xspace}

\usepackage{pifont}
\mathchardef\mhyphen="2D

\newfloat{lstfloat}{htbp}{lop}
\floatname{lstfloat}{Listing}

\usepackage{epsdice}
\usepackage{soul}
\usepackage[noabbrev,capitalize]{cleveref}

\usepackage{enumitem}

\newcommand{\ignore}[1]{}

\definecolor{TolDarkPurple}{HTML}{332288}
\definecolor{TolDarkBlue}{HTML}{6699CC}
\definecolor{TolLightBlue}{HTML}{88CCEE}
\definecolor{TolLightGreen}{HTML}{44AA99}
\definecolor{TolDarkGreen}{HTML}{117733}
\definecolor{TolDarkBrown}{HTML}{999933}
\definecolor{TolLightBrown}{HTML}{DDCC77}
\definecolor{TolDarkRed}{HTML}{661100}
\definecolor{TolLightRed}{HTML}{CC6677}
\definecolor{TolLightPink}{HTML}{AA4466}
\definecolor{TolDarkPink}{HTML}{882255}
\definecolor{TolLightPurple}{HTML}{AA4499}

\definecolor{PlotColorBlue}{HTML}{2C7FB8}
\definecolor{PlotColorRed}{HTML}{F03B20}
\definecolor{PlotColorGreen}{HTML}{31A354}

\definecolor{red}{HTML}{F03B20}
\definecolor{yellow}{HTML}{F5EE9A}
\definecolor{green}{HTML}{BEDB39}
\definecolor{blue}{HTML}{2C7FB8}

\pgfplotscreateplotcyclelist{mbarplot cycle}{%
  {draw=TolDarkBlue,    fill=TolDarkBlue!70},
  {draw=TolLightBrown,  fill=TolLightBrown!70},
  {draw=TolLightGreen,  fill=TolLightGreen!70},
  {draw=TolDarkPink,    fill=TolDarkPink!70},
  {draw=TolDarkPurple,  fill=TolDarkPurple!70},
  {draw=TolDarkRed,     fill=TolDarkRed!70},
  {draw=TolDarkBrown,   fill=TolDarkBrown!70},
  {draw=TolLightRed,    fill=TolLightRed!70},
  {draw=TolLightPink,   fill=TolLightPink!70},
  {draw=TolLightPurple, fill=TolLightPurple!70},
  {draw=TolLightBlue,   fill=TolLightBlue!70},
  {draw=TolDarkGreen,   fill=TolDarkGreen!70},
}

\pgfplotscreateplotcyclelist{mlineplot cycle}{%
  {TolDarkBlue, mark=*, mark size=1.5pt},
  {TolLightBrown, mark=square*, mark size=1.3pt},
  {TolLightGreen, mark=triangle*, mark size=1.5pt},
  {TolDarkBrown, mark=diamond*, mark size=1.5pt},
}

\pgfplotsset{
  compat=1.9
}
\pgfplotsset{
  mlineplot/.style={
  },
}

\pgfplotsset{
  mbarplot base/.style={
    mbaseplot,
    bar width=6pt,
    axis y line*=none,
  },
}

\pgfplotsset{
  mbarplot/.style={
    mbarplot base,
    ybar,
    xmajorgrids=false,
    ymajorgrids=true,
    area legend,
    legend image code/.code={%
      \draw[#1] (0cm,-0.1cm) rectangle (0.15cm,0.1cm);
    },
    cycle list name=mbarplot cycle,
  },
}

\pgfplotsset{
  horizontal mbarplot/.style={
    mbarplot base,
    xmajorgrids=true,
    ymajorgrids=false,
    xbar stacked,
    area legend,
    legend image code/.code={%
      \draw[#1] (0cm,-0.1cm) rectangle (0.15cm,0.1cm);
    },
    cycle list name=mbarplot cycle,
  },
}

\pgfplotsset{
  mbaseplot/.style={
    legend style={
      draw=none,
      fill=none,
      cells={anchor=west},
    },
    x tick label style={
      font=\footnotesize
    },
    y tick label style={
      font=\footnotesize
    },
    legend style={
      font=\footnotesize
    },
    major grid style={
      dotted,
    },
    axis x line*=bottom,
  },
  disable thousands separator/.style={
    /pgf/number format/.cd,
      1000 sep={}
  },
}

\newcommand{\ValFRCycle}{298}
\newcommand{\ValMinFRDetect}{596}

\newcommand{\ValDropItOverhead}{0.8}

\DeclareMathAlphabet\mathbfcal{OMS}{cmsy}{b}{n}
\newcommand{\POne}{$\mathbfcal{P}\mathbf{1}$\xspace}
\newcommand{\PTwo}{$\mathbfcal{P}\mathbf{2}$\xspace}
\newcommand{\PThree}{$\mathbfcal{P}\mathbf{3}$\xspace}

\setcopyright{rightsretained}
\acmDOI{XX.XXX/XXX_X}
\acmISBN{XXX-XXXX-XX-XXX/XX/XX}
\acmConference[ASIACCS'18]{ACM Asia Conference On Computer \& Communications Security}{June 4-8, 2018}{Songdo, Incheon, Korea}
\acmYear{2018}
\copyrightyear{2017}
\acmArticle{X}
\acmPrice{XX.XX}
\editor{Firstname Lastname}

\begin{document}
\sloppy                         

\title{Automated Detection, Exploitation, and Elimination of Double-Fetch Bugs using Modern CPU Features}
\author{
Michael Schwarz$^1$, Daniel Gruss$^1$, Moritz Lipp$^1$, Clémentine Maurice$^2$,\\Thomas Schuster$^1$, Anders Fogh$^3$, Stefan Mangard$^1$}
\affiliation{
$^1$ Graz University of Technology, Austria\\$^2$ CNRS, IRISA, France \\$^3$ G DATA Advanced Analytics, Germany
}

\begin{abstract}
Double-fetch bugs are a special type of race condition, where an unprivileged execution thread is able to change a memory location between the time-of-check and time-of-use of a privileged execution thread. 
If an unprivileged attacker changes the value at the right time, the privileged operation becomes inconsistent, leading to a change in control flow, and thus an escalation of privileges for the attacker. 
More severely, such double-fetch bugs can be introduced by the compiler, entirely invisible on the source-code level.

We propose novel techniques to efficiently detect, exploit, and eliminate double-fetch bugs.
We demonstrate the first combination of state-of-the-art cache attacks with kernel-fuzzing techniques to allow fully automated identification of double fetches.
We demonstrate the first fully automated reliable detection and exploitation of double-fetch bugs, making manual analysis as in previous work superfluous.
We show that cache-based triggers outperform state-of-the-art exploitation techniques significantly, leading to an exploitation success rate of up to 97\,\%. 
Our modified fuzzer automatically detects double fetches and automatically narrows down this candidate set for double-fetch bugs to the exploitable ones.
We present the first generic technique based on hardware transactional memory, to eliminate double-fetch bugs in a fully automated and transparent manner.
We extend defensive programming techniques by retrofitting arbitrary code with automated double-fetch prevention, both in trusted execution environments as well as in syscalls, with a performance overhead below 1\,\%.
\end{abstract}


\maketitle

\section{Introduction}
The security of modern computer systems relies fundamentally on the security of the operating system kernel, providing strong isolation between processes. 
While kernels are increasingly hardened against various types of memory corruption attacks, race conditions are still a non-trivial problem. 
Syscalls are a common scenario in which the trusted kernel space has to interact with the untrusted user space, requiring sharing of memory locations between the two environments. 
Among possible bugs in this scenario are time-of-check-to-time-of-use bugs, where the kernel accesses a memory location twice, first to check the validity of the data and second to use it (double fetch)~\cite{Serna2008}. 
If such double fetches are exploitable, they are considered double-fetch bugs.
The untrusted user space application can change the value between the two accesses and thus corrupt kernel memory and consequently escalate privileges.
Double-fetch bugs can not only be introduced at the source-code level but also by compilers, entirely invisible for the programmer and any source-code-level analysis technique~\cite{Blanchou2013}.
Recent research has found a significant amount of double fetches in the kernel through static analysis~\cite{Wang2017}, and memory access tracing through full CPU emulation~\cite{Jurczyk2013}.
Both works had to manually determine for every double fetch, whether it is a double-fetch bug.

Double fetches have the property that the data is fetched twice from memory. 
If the data is already in the cache (\emph{cache hit}), the data is fetched from the cache, if the data is not in the cache (\emph{cache miss}), it is fetched from main memory into the cache. 
Differences between fetches from cache and memory are the basis for so-called cache attacks, such as \FlushReload \cite{Osvik2006,Yarom2014}, which obtain secret information by observing memory accesses~\cite{Ge2016}. 
Instead of exploiting the cache side channel for obtaining secret information, we utilize it to detect double fetches.

In this paper, we show how to efficiently and automatically detect, exploit, and eliminate double-fetch bugs, with two new approaches: DECAF and \DropIt.

DECAF is a double-fetch-exposing cache-guided augmentation for fuzzers, which automatically \emph{detects} and \emph{exploits} real-world double-fetch bugs in a two-phase process.
In the profiling phase, DECAF relies on cache side channel information to detect whenever the kernel accesses a syscall parameter. 
Using this novel technique, DECAF is able to detect whether a parameter is fetched multiple times, generating a candidate set containing double fetches, \ie some of which are potential double-fetch bugs.
In the exploitation phase, DECAF uses a cache-based trigger signal to flip values while fuzzing syscalls from the candidate set, to trigger actual double-fetch bugs.
In contrast to previous purely probability-based approaches, cache-based trigger signals enable deterministic double-fetch-bug exploitation.
Our automated exploitation exceeds state-of-the-art techniques, where checking the double-fetch candidate set for actual double-fetch bugs is tedious manual work.
We show that DECAF can also be applied to trusted execution environments, \eg ARM TrustZone and Intel SGX.

\DropIt is a protection mechanism to \emph{eliminate} double-fetch bugs.
\DropIt uses hardware transactional memory to efficiently drop the current execution state in case of a concurrent modification.
Hence, double-fetch bugs are automatically reduced to ordinary non-exploitable double fetches.
In case user-controlled memory locations are modified, \DropIt continues the execution from the last consistent state.
Applying \DropIt to syscalls induces no performance overhead on arbitrary computations running in other threads and only a negligible performance overhead of \SI{\ValDropItOverhead}{\percent} on the process executing the protected syscall.
We show that \DropIt can also be applied to trusted execution environments, \eg ARM TrustZone and Intel SGX. 

\subheading{Contributions.} We make the following contributions:
\begin{compactenum}
  \item We are the first to combine state-of-the-art cache attacks with kernel-fuzzing techniques to build DECAF, a generic double-fetch-exposing cache-guided augmentation for fuzzers.
  \item Using DECAF, we are the first to show fully automated reliable detection and exploitation of double-fetch bugs, making manual analysis as in previous work superfluous.
  \item We outperform state-of-the-art exploitation techniques significantly, with an exploitation success rate of up to 97\,\%. 
  \item We present \DropIt, the first generic technique to eliminate double-fetch bugs in a fully automated manner, facilitating newfound effects of hardware transactional memory on double-fetch bugs. \DropIt has a negligible performance overhead of \SI{\ValDropItOverhead}{\percent} on protected syscalls.
  \item We show that DECAF can also fuzz trusted execution environments in a fully automated manner. We observe strong synergies between Intel SGX and \DropIt, enabling efficient preventative protection from double-fetch bugs.
\end{compactenum}

\subheading{Outline.}
The remainder of the paper is organized as follows. 
In Section~\ref{sec:background}, we provide background on cache attacks, race conditions, and kernel fuzzing.
In Section~\ref{sec:framework}, we discuss the building blocks for finding and eliminating double-fetch bugs. 
We present the profiling phase of DECAF in Section~\ref{sec:discovery} and the exploitation phase of DECAF in Section~\ref{sec:exploitation}.
In Section~\ref{sec:prevention}, we show how hardware transactional memory can be used to eliminate all double-fetch bugs generically.
In Section~\ref{sec:evaluation} we discuss the results of we obtained by instantiating DECAF. 
We conclude in Section~\ref{sec:conclusion}.

\section{Background}\label{sec:background}

\subsection{Fuzzing}

Fuzzing describes the process of testing applications with randomized input to find vulnerabilities. 
Due to the cost-effectiveness and good results, Duran and Ntafos~\cite{Duran1981} described fuzzing as a viable strategy when testing applications.

The term ``fuzzing'' was coined 1988 by Miller~\cite{Miller1990} who tested the effects of noise over ``fuzzy'' network connections on UNIX applications.
This work was extended to an automated approach for testing the reliability of several user-space programs on Linux~\cite{Miller1995}, Windows~\cite{Forrester2000} and Mac OS~\cite{Miller2006}.
There is an immense number of works exploring user space fuzzing with different forms of feedback~\cite{Godefroid2005,Godefroid2007,Godefroid2008,Eddington2008,Ganesh2009,Godefroid2012,Haller2013,Copos2015,Kargen2015,Stephens2016,Rawat2017}.
However, these are not applicable to this work, as we focus on fuzzing the kernel and trusted execution environments.

Fuzzing is not limited to testing user-space applications, but it is also, to a much smaller extent, used to test the reliability of operating systems. 
Regular user space fuzzers cannot be used here, but a smaller number of tools have been developed to apply fuzzy testing to operating system interfaces. 
Carrette~\cite{Carrette1996} developed the tool CrashMe that tests the robustness of operating systems by trying to execute random data streams as instructions.
Mendoncca~\etal\cite{Mendoncca2008} and Jodeit~\etal\cite{Jodeit2010} demonstrate that fuzzing drivers via the hardware level is another possibility to attack an operating system. 
Other operating system interfaces that can be fuzzed include the file system~\cite{Cadar2008} and the virtual machine interface~\cite{Martignoni2010,Gauthier2011}.

The syscall interface is a trust boundary between the trusted kernel, running with the highest privileges, and the unprivileged user space. 
Bugs in this interface can be exploited to escalate privileges.
Koopman~\etal\cite{Koopman1997} were among the first to test random inputs to syscalls. 
Modern syscall fuzzers, such as Trinity~\cite{Jones2011} or syzkaller~\cite{Vyukov2016}, test most syscalls with semi-intelligent arguments instead of totally random inputs. 
In contrast to these generic tools, Weaver~\etal\cite{Weaver2015} developed perf\_fuzzer, which uses domain knowledge to fuzz only the performance monitoring syscalls. 

\subsection{\FlushReload}

\FlushReload is a side-channel attack exploiting the difference in access times between CPU caches and main memory. 
Yarom and Falkner~\cite{Yarom2014} presented \FlushReload as an improvement over the cache attack by Gullasch~\etal\cite{Gullasch2011}.
\FlushReload relies on shared memory between the attacker and the victim and works as follows:
\begin{compactenum}
 \item Establish a shared memory region with the victim (\eg by mapping the victim binary into the address space).
 \item Flush one line of the shared memory from the cache. 
 \item Schedule the victim process.
 \item Measure the access time to the flushed cache line.
\end{compactenum}
If the victim accesses the cache line while being scheduled, it is again cached. 
When measuring the access time, the attacker can distinguish whether the data is cached or not and thus infer whether the victim accessed it. 
As \FlushReload works on cache line granularity (usually \SI{64}{B}), fine-grained attacks are possible. 
The probability of false positives is very low with \FlushReload, as cache hits cannot be caused by different programs and prefetching can be avoided.
Gruss~\etal\cite{Gruss2016Flush} reported extraordinarily high accuracies, above \SI{99}{\percent}, for the \FlushReload side channel, making it a viable choice for a wide range of applications.

\subsection{Double Fetches and Double-Fetch Bugs}\label{sec:dfadfb}
In a scenario where shared memory is accessed multiple times, the CPU may fetch it multiple times into a register.
This is commonly known as a \textbf{double fetch}.
Double fetches occur when the kernel accesses data provided by the user multiple times, which is often unavoidable. 
If proper checks are done, ensuring that a change in the data during the fetches is correctly handled, double fetches are \textbf{non-exploitable} valid constructs.

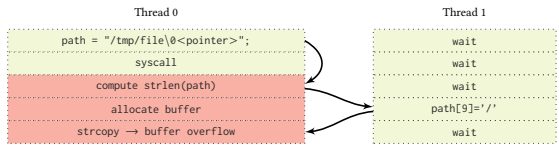
\begin{figure}[t]
 \centering
 \resizebox{0.9\hsize}{!}{
 \resizebox{\hsize}{!}{
\begin{tikzpicture}[scale=0.4]
 \tikzstyle{every node}=[font=\scriptsize]
		\draw (6.5,5.75) node[text centered] {Thread 0};

		\draw[color=black!85!white,fill=green!20, dotted] (0,4) rectangle +(13,1) node[midway] {\texttt{path = "/tmp/file$\backslash$0$<$pointer$>$"$\texttt{;}$}}; 
		\draw[color=black!85!white,fill=green!20, dotted] (0,3) rectangle +(13,1) node[midway] {\texttt{syscall}}; 
		\draw[color=black!85!white,fill=red!40, dotted] (0,2) rectangle +(13,1) node[midway] {\texttt{compute strlen(path)}}; 
		\draw[color=black!85!white,fill=red!40, dotted] (0,1) rectangle +(13,1) node[midway] {\texttt{allocate buffer}}; 
		\draw[color=black!85!white,fill=red!40, dotted] (0,0) rectangle +(13,1) node[midway] {\texttt{strcopy $\to$ buffer overflow}}; 

\begin{scope}[shift={(16,0)}]
		\draw (4,5.75) node[text centered] {Thread 1};

		\draw[color=black!85!white,fill=green!20, dotted] (0,4) rectangle +(8,1) node[midway] {\texttt{wait}}; 
		\draw[color=black!85!white,fill=green!20, dotted] (0,3) rectangle +(8,1) node[midway] {\texttt{wait}}; 
		\draw[color=black!85!white,fill=green!20, dotted] (0,2) rectangle +(8,1) node[midway] {\texttt{wait}}; 
		\draw[color=black!85!white,fill=green!20, dotted] (0,1) rectangle +(8,1) node[midway] {\texttt{path[9]='/'}}; 
		\draw[color=black!85!white,fill=green!20, dotted] (0,0) rectangle +(8,1) node[midway] {\texttt{wait}}; 
\end{scope}

\draw[-latex',thick,out=-30,in=30,looseness = 1.5] (13,4.5) to (13,2.6);
\draw[-latex',thick,out=-5,in=175] (13,2.4) to (16,1.6);
\draw[-latex',thick,out=185,in=5] (16,1.4) to (13,0.5);

		\end{tikzpicture} 
}
 }
 \caption{A double-fetch bug exploited from a second thread by replacing the 0-byte after the length check.}
 \label{fig:double-fetch-explanation}
\end{figure}

A \textbf{double-fetch bug} is a time-of-check-to-time-of-use race condition, which is \textbf{exploitable} by changing the data in the shared memory between two accesses.
Double-fetch bugs are a subset of double fetches.
A double fetch is a double-fetch bug, if and only if it can be exploited by concurrent modification of the data.
For example, if a syscall expects a string and first checks the length of the string before copying it to the kernel, an attacker could change the string to a longer string after the check, causing a buffer overflow in the kernel (Figure~\ref{fig:double-fetch-explanation}).
In the worst case, this leads to code execution within the kernel.

Wang~\etal\cite{Wang2017} used Coccinelle, a transformation and matching engine for C code, to find double fetches.
With this static pattern-based approach, they identified \SIx{90} double fetches inside the Linux kernel.
However, their work incurred several days of manual analysis of these \SIx{90} double fetches, identifying only \SIx{3} exploitable double-fetch bugs.
A further limitation of their work is that double-fetch bugs which do not match the implemented patterns, cannot be detected.

Not all double fetches, and thus not all double-fetch bugs, can be found using static code analysis. 
Blanchou~\cite{Blanchou2013} demonstrated that especially in lock-free code, compilers can introduce double fetches that are not present in the code. 
Even worse, these compiler-introduced double fetches can become double-fetch bugs in certain scenarios (\eg CVE-2015-8550). 
Jurczyk~\etal\cite{Jurczyk2013} presented a dynamic approach for finding double fetches.
They used a full CPU emulator to run Windows and log all memory accesses.
Note that this requires significant computation and storage resources, as just booting Windows already consumes 15 hours of time, resulting in a log file of more than \SI{100}{GB}~\cite{Jurczyk2013}.
In the memory access log, they searched for a distinctive double-fetch pattern, \eg two reads of the same user-space address within a short time frame. 
They identified \SIx{89} double fetches in Windows 7 and Windows 8.
However, their work also required manual analysis, in which they found that only 2 out of around 100 unique double fetches were exploitable double-fetch bugs.
Again, if a double-fetch bug does not match the implemented double-fetch pattern, it is not detected.
In summary, we find that all techniques for double-fetch bug detection are probabilistic and hence incomplete.

\subsubsection{Race Condition Detection}
Besides research on double fetches and double-fetch bugs, there has been a significant amount of research on race condition detection in general.
Static analysis of source code and dynamic runtime analysis have been used to find data race conditions in multithreaded applications.
Savage~\etal\cite{Savage1997} described the Lockset algorithm. Their tool, Eraser, dynamically detects race conditions in multithreaded programs.
Pozniansky~\etal\cite{Pozniansky2003, Pozniansky2007} extended their work to detect race conditions in multithreaded C++ programs on-the-fly.
Yu~\etal\cite{Yu2005} described RaceTrack, an adaptive detection algorithm that reports suspicious activity patterns.
These algorithms have been improved and made more efficient by using more lightweight data structures~\cite{Flanagan2010} or combining various approaches~\cite{Xie2011}.

While these tools can be applied to user space programs, they are not designed to detect race conditions in the kernel space.
Erickson~\etal\cite{Erickson2010} utilized breakpoints and watchpoints on memory accesses to detect data races in the Windows kernel.
With RaceHound~\cite{RaceHound}, the same idea has been implemented for the Linux kernel.
The SLAM~\cite{slam} project uses symbolic model checking, program analysis, and theorem proving, to verify whether a driver correctly interacts with the operating system.
Schwarz~\etal\cite{Schwarz2005} utilized software model checking to detect security violations in a Linux distribution.

More closely related to double-fetch bugs, other time-of-check-to-time-of-use bugs exist.
By changing the content of a memory location that is passed to the operating system, the content of a file could be altered after a validity check~\cite{Bishop1996race,Wei2005,Cai2009}. 
Especially time-of-check-to-time-of-use bugs in the file system are well-studied, and several solutions have been proposed~\cite{Cowan2001raceguard,Uppuluri2005,Lhee2005racecondition,Payer2012tocttou}. 

\subsection{Hardware Transactional Memory}\label{sec:tsx}
Hardware transactional memory is designed for optimizing synchronization primitives~\cite{Yoo2013,Ferri2009}.
Any changes performed inside a transaction are not visible to the outside before the transaction succeeds.
The processor speculatively lets a thread perform a sequence of operations inside a transaction.
Unless there is a conflict due to a concurrent modification of a data value, the transaction succeeds.
However, if a conflict occurs before the transaction is completed (\eg a concurrent write access), the transaction aborts.
In this case, all changes that have been performed in the transaction are discarded, and the previous state is recovered.
These fundamental properties of hardware transactional memory imply that once a value is read in a transaction, the value cannot be changed from outside the transaction anymore for the time of the transaction.

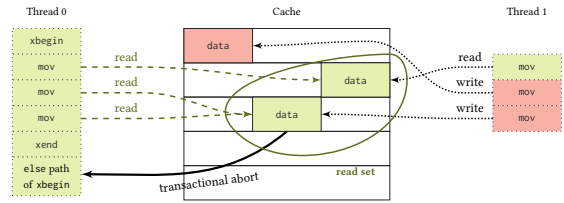
\begin{figure}[t]
 \centering
 \resizebox{0.9\hsize}{!}{
 \resizebox{\hsize}{!}{
\begin{tikzpicture}[scale=0.65]
 \tikzstyle{every node}=[font=\scriptsize]

		\draw (17,6.5) node[text centered] {Thread 1};

		\draw (3,6.5) node[text centered] {Thread 0};
		
		\draw (10,6.5) node[text centered] {Cache};
		\draw[color=black!85!white] (7,1) rectangle +(6,1); 
		\draw[color=black!85!white] (7,2) rectangle +(6,1); 
		\draw[color=black!85!white] (7,3) rectangle +(6,1); 
		\draw[color=black!85!white] (7,4) rectangle +(6,1); 
		\draw[color=black!85!white] (7,5) rectangle +(6,1); 

\begin{scope}[yscale=0.75,shift={(0,2)}]
		\draw[color=black!85!white,fill=green!40, dotted] (16,4) rectangle +(2,1) node[midway] {\texttt{mov}}; 
		\draw[color=black!85!white,fill=red!40, dotted] (16,3) rectangle +(2,1) node[midway] {\texttt{mov}}; 
		\draw[color=black!85!white,fill=red!40, dotted] (16,2) rectangle +(2,1) node[midway] {\texttt{mov}}; 
		\draw[color=black!85!white,fill=green!40, dotted] (2,5) rectangle +(2,1) node[midway] {\texttt{xbegin}}; 
		\draw[color=black!85!white,fill=green!40, dotted] (2,4) rectangle +(2,1) node[midway] {\texttt{mov}}; 
		\draw[color=black!85!white,fill=green!40, dotted] (2,3) rectangle +(2,1) node[midway] {\texttt{mov}}; 
		\draw[color=black!85!white,fill=green!40, dotted] (2,2) rectangle +(2,1) node[midway] {\texttt{mov}}; 
		\draw[color=black!85!white,fill=green!40, dotted] (2,1) rectangle +(2,1) node[midway] {\texttt{xend}}; 
		\draw[color=black!85!white,fill=green!40, dotted] (2,-0.5) rectangle +(2,1.5);
		\draw (3,0.5) node {\scriptsize \texttt{else} path}; 
		\draw (3,-0.1) node {\scriptsize of \texttt{xbegin}}; 
\end{scope}

		\draw[color=black!85!white,fill=green!40] (9,3) rectangle +(2,1) node[midway] {\texttt{data}}; 
		\draw[-latex',thick,out=0,in=180,dashed,green!50!black] (4,4.875)  to node[near end,sloped,above]{\footnotesize read} +(1.8,0) to (11,4.5);
		
		\draw[-latex',thick,densely dotted,out=180,in=0] (16,4.875)  to node[near end,sloped,above]{\footnotesize read} +(-0.9,0) to (13,4.5);
		
		\draw[color=black!85!white,fill=green!40] (11,4) rectangle +(2,1) node[midway] {\texttt{data}}; 
		\draw[-latex',thick,out=0,in=180,dashed,green!50!black] (4,4.125) to node[near end,sloped,above]{\footnotesize read} +(1.8,0) to  (9,3.5);
				
		\draw[color=black!85!white,fill=red!40] (7,5) rectangle +(2,1) node[midway] {\texttt{data}}; 
		\draw[-latex',thick,densely dotted,out=180,in=0] (16,4.125) to node[near end,sloped,above]{\footnotesize write} +(-0.9,0) to +(-2.7,+1.5) to (9,5.5);
		
		\draw[-latex',thick,out=0,in=180,dashed,green!50!black] (4,3.375)  to node[near end,sloped,above]{\footnotesize read} +(1.8,0) to (9,3.5);
		\draw[-latex',thick,densely dotted,out=180,in=0] (16,3.375) to node[near end,sloped,above]{\footnotesize write} +(-0.9,0) to (11,3.5);
		
				\draw[-latex',very thick,out=-140,in=0] (10,3) to node[midway,sloped,below,xshift=1em]{\footnotesize transactional abort} (4,1.75);
				
    \draw[green!50!black,thick,in=90,out=0] (13,5.25) to (13.5,4.5);
    \draw[green!50!black,thick,in=-45,out=-90] (13.5,4.5) to (8.5,3);
    \draw (12,1.85) node {\textcolor{green!50!black}{\textbf{read set}}};
    \draw[green!50!black,thick,in=180,out=135] (8.5,3) to (13,5.25);

		\end{tikzpicture} 
}
 }
 \caption{Memory locations are automatically added to read and write set. Upon conflicting memory accesses a transactional abort unrolls all transactional operations and jumps to the else path of \texttt{xbegin}.}
 \label{fig:htm}
\end{figure}

Intel TSX implements hardware transactional memory on a cache line granularity.
It maintains a read set which is limited to the size of the L3 cache and a write set which is limited to the size of the L1 cache~\cite{Intel_opt,Goel2014,Liu2014concurrent,Zacharopoulos2015}.
A cache line is automatically added to the read set when it is read inside a transaction, and it is automatically added to the write set when it is modified inside a transaction.
Modifications to any memory in the read set or write set from other threads cause the transaction to abort, as illustrated in Figure~\ref{fig:htm}.

Previous work has investigated whether hardware transactional memory can be instrumented for security features.
Guan~\etal\cite{Guan2015} proposed to protect cryptographic keys by only decrypting them within TSX transactions.
As the keys are never written to DRAM in an unencrypted form, they cannot be read from memory even by a physical attacker probing the DRAM bus.
Kuvaiskii~\etal\cite{Kuvaiskii2016} proposed to use TSX to detect hardware faults and roll-back the system state in case a fault occurred.
Shih~\etal\cite{Shih2017tsgx} proposed to exploit the fact that TSX transactions abort if a page fault occurred for a memory access to prevent controlled-channel attacks~\cite{Xu2015controlled} in cloud scenarios.
Chen~\etal\cite{Chen2017} implemented a counting thread protected by TSX to detect controlled-channel attacks in SGX enclaves.
Gruss~\etal\cite{Gruss2017Cloak} demonstrated that TSX can be used to protect against cache side-channel attacks in the cloud.

Shih~\etal\cite{Shih2017tsgx} and Gruss~\etal\cite{Gruss2017Cloak} observed that Intel TSX has several practical limitations.
One observation is that executed code is not considered transactional memory, \ie virtually unlimited amount of code can be executed in a transaction.
To evade the limitations caused by the L1 and L3 cache sizes, Shih~\etal\cite{Shih2017tsgx} and Gruss~\etal\cite{Gruss2017Cloak} split transactions that might be memory-intense into multiple smaller transactions.

\section{Building Blocks to Detect, Exploit, and Eliminate Double-Fetch Bugs}\label{sec:framework}

In this section, we present building blocks for detecting double fetches, exploiting double-fetch bugs, and eliminating double-fetch bugs.
These building blocks are the base for DECAF and \DropIt.

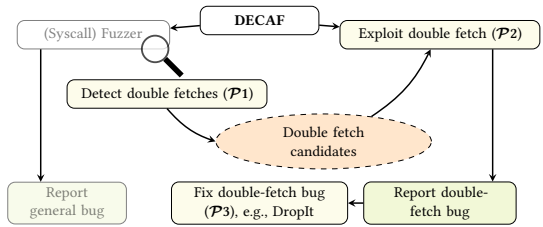
\begin{figure}[t]
 \centering
 \tikzstyle{process} = [rectangle, rounded corners, minimum width=2.5cm, minimum height=.6cm,text centered, draw=black, fill=white]
\tikzstyle{round} = [ellipse, rounded corners, minimum width=2cm, minimum height=.6cm,text centered, draw=black,dashed, fill=white]
\tikzstyle{processdis} = [rectangle, rounded corners, minimum width=2.5cm, minimum height=.6cm,text centered, draw=black!60!white, fill=white!90!black,text=black!60!white]
\tikzstyle{thread} = [rectangle, minimum width=2cm, minimum height=.6cm, text centered, draw=black, fill=orange!30]
\tikzstyle{arrow} = [thick,->,>=stealth]
\tikzstyle{arrowdis} = [thick,->,>=stealth,draw=black!60!white,fill=black!60!white]

\usetikzlibrary{shapes.geometric, arrows}

\resizebox{0.9\hsize}{!}{
\begin{tikzpicture}[node distance=0.3cm,scale=0.9,transform shape]
\draw [draw=none] (-6,-1) -- (-6,-1); 

\node (trinity) [process] {\textbf{DECAF}};
\node (trinitymain) [process, below of=trinity, xshift=-3.5cm,text width=3cm,opacity=0.5] {(Syscall) Fuzzer};
\node (exploit) [process, right of=trinitymain, xshift=7cm,text width=4cm,fill=yellow!20] {Exploit double fetch (\PTwo)};
\begin{scope}[transform canvas={xshift = -0.5cm}]
\node (child1) [process, below of=trinitymain, xshift=0cm,yshift=-3.25cm,fill=green!20,text width=2cm,opacity=0.5] {Report general bug};
\end{scope}
\node (monitor) [process, right of=trinitymain, xshift=1.25cm,yshift=-1.25cm,text width=4cm,fill=yellow!20] {Detect double fetches (\POne)};
\node (doublefetches) [round, right of=monitor,text width=3cm,xshift=3cm,yshift=-1cm,fill=orange!20] {Double fetch candidates};
\node (reportdf) [process, below of=exploit, xshift=0cm,yshift=-3.25cm,fill=green!20,text width=3cm] {Report double-fetch bug};
\node (dropit) [process, left of=reportdf, xshift=-3.5cm,yshift=0cm,text width=3.5cm,fill=yellow!20] {Fix double-fetch bug (\PThree), \eg \DropIt};

\begin{scope}[shift={(-8.5cm,-2.0cm)},scale=0.7]
\draw [draw=black,line width=3pt] (9.3,1.7) to (9.8,1.2);
\draw [draw=black,line width=1pt,fill=white!10,opacity=0.7] (9,2) ellipse (0.4cm and 0.4cm);
\end{scope}

\draw [arrow] (trinity) -- (trinitymain);
\draw [arrow] (trinity) -- (exploit);
\begin{scope}[transform canvas={xshift = -1cm}]
\draw [arrow] (trinitymain) -- (child1);
\end{scope}
\draw [arrow,bend right=15] (doublefetches) to node[below,sloped]{} (exploit);
\begin{scope}[transform canvas={xshift = 1cm}]
\draw [arrow] (exploit) -- (reportdf)  ;
\end{scope}
\draw [arrow] (reportdf) -- (dropit);

\draw [arrow, bend right=15] (monitor.south) to node[below,sloped] {} (doublefetches.west);

\end{tikzpicture}
}
 \caption{An overview of the framework. Detecting (Primitive \POne) and exploiting (Primitive \PTwo) double fetches runs in parallel to the syscall fuzzer. Reported double-fetch bugs can be eliminated (Primitive \PThree) after the fuzzing process.}
 \label{fig:framework}
\end{figure}

We identified three primitives, illustrated in~\Cref{fig:framework}, for which we propose novel techniques in this paper:
\begin{compactenum}
 \item[\POne:] Detecting double fetches by using the \FlushReload side channel.
 \item[\PTwo:] Distinguishing (exploitable) double-fetch bugs from (non-exploitable) double fetches by validating their exploitability by automatically exploiting double-fetch bugs.
 \item[\PThree:] Eliminating (exploitable) double-fetch bugs by using hardware transactional memory.
\end{compactenum}

In \Cref{sec:discovery}, we propose a novel, fully automated technique to detect double fetches (\POne) using a multi-threaded \FlushReload cache side-channel attack. 
Our technique complements other work on double-fetch bug detection~\cite{Jurczyk2013,Wang2017} as it covers scenarios which lead to false positives and false negatives in other detection methods.
Although relying on a side channel may seem unusual, this approach has certain advantages over state-of-the-art techniques, such as memory access tracing~\cite{Jurczyk2013} or static code analysis~\cite{Wang2017}. 
We do not need any model of what constitutes a double fetch in terms of memory access traces or static code patterns.
Hence, we can detect any double fetch regardless of any double fetch model.

Wang~\etal\cite{Wang2017} identified as limitations of their initial approach that false positives occur if a pointer is changed between two fetches and memory accesses, in fact, go to different locations or if user-space fetches occur in loops.
Furthermore, false negatives occur if multiple pointers point to the same memory location (pointer aliasing) or if memory is addressed through different types (type conversion), or if an element is fetched separately from the corresponding pointer and memory.
With a refined approach, they reduced the false positive rate from more than 98\,\% to only 94\,\%, \ie 6\,\% of the detected situations turned out to be actual double-fetch bugs in the manual analysis.
Wang~\etal\cite{Wang2017} reported that it took an expert ``only a few days'' to analyze them.
In contrast, our \FlushReload-based approach is oblivious to language-level structures.
The \FlushReload-trigger only depends on actual accesses to the same memory location, irrespective of any programming constructs.
Hence, we inherently bypass the problems of the approach of Wang~\etal\cite{Wang2017} by design.

Our technique does not replace existing tools, which are either slow~\cite{Jurczyk2013} or limited by static code analysis~\cite{Wang2017} and require manual analysis.
Instead, we complement previous approaches by utilizing a side channel, allowing fully automatic detection of double-fetch bugs, including those that previous approaches may miss.

In \Cref{sec:exploitation}, we propose a novel technique to automatically determine whether a double fetch found using \POne is an (exploitable) double-fetch bug (\PTwo).
State-of-the-art techniques are only capable of automatically detecting double fetches using either dynamic~\cite{Jurczyk2013} or static~\cite{Wang2017} code analysis, but cannot determine whether a found double fetch is an exploitable double-fetch bug. 
The double fetches found using these techniques still require manual analysis to check whether they are valid constructs or exploitable double-fetch bugs.
We close this gap by automatically testing whether double fetches are exploitable double-fetch bugs (\PTwo), eliminating the need for manual analysis. 
Again, this technique relies on a cache side channel to trigger a change of the double-fetched value between the two fetches (\PTwo). 
This is not possible with previous techniques~\cite{Jurczyk2013,Wang2017}.

As the first automated technique, we present DECAF, a double-fetch-exposing cache-guided augmentation for fuzzers, leveraging \POne and \PTwo in parallel to regular fuzzing.
This allows us to automatically detect double fetches in the kernel and to automatically narrow them down to the exploitable double-fetch bugs (\cf \Cref{sec:exploitation}), as opposed to previous techniques~\cite{Jurczyk2013,Wang2017} which incurred several days of manual analysis work by an expert to distinguish double-fetch bugs from double fetches.
Similar to previous approaches~\cite{Jurczyk2013,Wang2017}, which inherently could not detect all double-fetch bugs in the analyzed code base, our approach is also probabilistic and might not detect all double-fetch bugs.
However, due to their different underlying techniques, the previous approaches and ours complement each other.

In \Cref{sec:prevention}, we present a novel method to simplify the elimination of detected double-fetch bugs (\PThree).
We observe previously unknown interactions between double-fetch bugs and hardware transactional memory.
Utilizing these effects, \PThree can protect code without requiring to identify the actual cause of a double-fetch bug.
Moreover, \PThree can even be applied as a preventative measure to protect critical code.

As a practical implementation of \PThree, we built \DropIt, an open-source~\footnote{The source can be found in an anonymous GitHub repository \href{https://www.github.com/libdropit/libdropit}{\texttt{https://www.github.com/libdropit/libdropit}}.} instantiation of \PThree based on Intel TSX.
We implemented \DropIt as a library, which eliminates double-fetch bugs with as few as 3 additional lines of code. 
We show that \DropIt has the same effect as rewriting the code to eliminate the double fetch. 
Furthermore, \DropIt can automatically and transparently eliminate double-fetch bugs in trusted execution environments such as Intel SGX, in both desktop and cloud environments.

\section{Detecting Double Fetches}\label{sec:discovery}

We propose a novel dynamic approach to detect double fetches based on their cache access pattern (\POne, \cf \Cref{sec:framework}). 
The main idea is to monitor the cache access pattern of syscall arguments of a certain type, \ie pointers or structures containing pointers.
These pointers may be accessed multiple times by the kernel and, hence, a second thread can change the content.
Other arguments that are statically copied or passed by value, and consequently are not accessed multiple times, cannot lead to double fetches.

To monitor access to potentially vulnerable function arguments, we mount a \FlushReload attack on each argument in dedicated monitoring threads.
A monitoring thread continuously flushes and reloads the memory location referenced by the function argument. 
As soon as the kernel accesses the function argument, the data is loaded into the cache.
In this case, the \FlushReload attack in the corresponding monitoring thread reports a \emph{cache hit}.

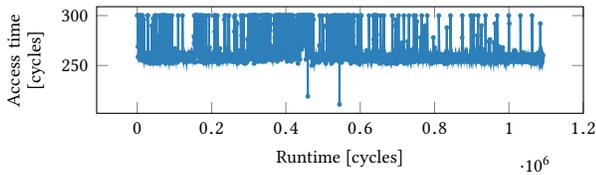
\begin{figure}[t]
 \centering
 \begin{tikzpicture}
\begin{axis}[
mlineplot,
style={font=\footnotesize},
xlabel={Runtime [cycles]},
ylabel={Access time [cycles]},
ylabel style={text width=2cm,align=center},
width=0.95\hsize,
scaled y ticks=false,
height=3cm
]
\addplot+[blue,thick,mark options={draw=blue,fill=blue},mark=*,mark size=0.5] table[x=Time,y=Access,col sep=space] {data/out_2.csv};

\end{axis}
\end{tikzpicture}
 \caption{\FlushReload timing trace for a syscall with a double fetch. The two downward peaks show when the kernel accessed the argument.}
 \label{fig:cache-hit-2}
\end{figure}

Figure~\ref{fig:cache-hit-2} shows a trace generated by a monitoring thread. 
The trace contains the access time in cycles for the memory location referenced by the function argument. 
If the memory is accessed twice, \ie a double fetch, we can see a second cache hit, as shown in Figure~\ref{fig:cache-hit-2}. 
This provides us with primitive \POne.

\subsection{Classification of Multiple Cache Hits}\label{sec:classification}

Multiple cache hits within one trace usually correspond to multiple fetches.
However, there are rare cases where this is not the case.
To entirely eliminate spurious cache hits from prefetching, we simply disabled the prefetcher in software through MSR \texttt{0x1A4} and allocated memory on different pages to avoid spatial prefetching.
Note that this does not have any effect on the overall system stability and only a small performance impact.
We want to discuss two other factors influencing the cache access pattern in more detail.

\paragraph{Size of data type}
Depending on the size of the data type, there are differences in the cache access pattern. 
If the data fills exactly one cache line, accesses to the cache line are clearly seen in the cache access pattern. 
There are no false positives due to unrelated data in the same cache set, and every access to the referenced memory is visible in the cache access pattern. 

To avoid false positives if the data size is smaller than a cache line (\ie \SI{64}{B}), we allocate memory chunks with a multiple of the page size, ensuring that dynamically allocated memory never shares one cache line. 
Hence, accesses to unrelated data (\ie separate allocations) do not introduce any false positives, as they are never stored in the same cache line.
Thus, false positives are only detected if the cache line contains either multiple parameters, local variables or other members of the same structure. 

\paragraph{Parameter reuse}

With call-by-reference, one parameter of a function can be used both as input and output, \eg in functions working in-place on a given buffer. 
Using \FlushReload, we cannot distinguish whether a cache hit is due to a read of or write to the memory.
Thus, we can only observe multiple cache hits without knowing whether they are both caused by a memory read access or by other activity on the same cache line.

\subsection{Probability of Detecting a Double Fetch}\label{sec:prob_df_detection}

The actual detection rate of a double fetch depends on the time between two accesses. 
Each \FlushReload cycle consists of flushing the memory from the cache and measuring the access time to this memory location afterwards. 
Such a cycle takes on average \SIx{\ValFRCycle} cycles on an Intel i7-6700K.
Thus, to detect a double fetch, the time between the two memory accesses has to be at least two \FlushReload cycles, \ie \SIx{\ValMinFRDetect} CPU cycles. 

We obtain the exact same results when testing a double fetch in kernel space as in user space.
Also, due to the high noise-resistance of \FlushReload (\cf \Cref{sec:background}), interrupts, context switches, and other system activity have an entirely negligible effect on the result.
With the minimum distance of \SIx{\ValMinFRDetect} CPU cycles, we can already detect double fetches if the scheduling is optimal for both applications. 
The further the memory fetches are apart, the higher the probability of detecting the double fetch. 
The probability of detecting double fetches increases monotonically with the time between the fetches, making it quite immune to interrupts such as scheduling. 
If the double fetches are at least \SIx{3000} CPU cycles apart, we almost always detect such a double fetch. 
In the real-world double-fetch bugs we examined, the double fetches were always significantly more than \SIx{3000} CPU cycles apart.
\cref{fig:double-fetch} (\Cref{sec:appendix_dfetches}) shows the relation between the detection probability and the time between the memory accesses, empirically determined on an Intel i7-6700K.

On a Raspberry Pi 3 with an ARMv8 1.2\,GHz Broadcom BCM2837 CPU, a \FlushReload cycle takes \SIx{250} cycles on average.
Hence, the double fetches must be at least \SIx{500} cycles apart to be detectable with a high probability.

\subsection{Automatically Finding Affected Syscalls}\label{sec:trinitydecaf_syscalls}

Using our primitive \POne, we can already automatically and reliably detect whether a double fetch occurs for a particular function parameter. 
This is the first building block of DECAF.
DECAF is a two-phase process, consisting of a profiling phase which finds double fetches and an exploitation phase narrowing down the set of double fetches to only double-fetch bugs.
We will now discuss how DECAF augments existing fuzzers to discover double fetches within operating system kernels fully automatically. 

To test a wide range of syscalls and their parameters, we instantiate DECAF with existing syscall fuzzers. 
For Linux, we retrofitted the well-known syscall fuzzer \emph{Trinity} with our primitive \POne. 
For Windows, we extended the basic \emph{NtCall64} fuzzer to support semi-intelligent parameter selection similar to Trinity. 
Subsequently, we retrofitted our extended NtCall64 fuzzer with our primitive \POne as well.
Thereby, we demonstrate that DECAF is a generic technique and does not depend on a specific fuzzer or operating system.

Our augmented and extended NtCall64 fuzzer, \emph{NtCall64DECAF} works for double fetches and double-fetch bugs in proof-of-concept drivers.
However, due to the severely limited coverage of the NtCall64 fuzzer, we did not include it in our evaluations.
Instead, we focus on Linux only and leave retrofitting a good Windows syscall fuzzer with DECAF for future work. 

In the profiling phase of DECAF, the augmented syscall fuzzer chooses a random syscall to test. 
The semi-intelligent parameter selection of the syscall fuzzer ensures that the syscall parameters are valid parameters in most cases. 
Hence, the syscall is executed and does not abort in the initial sanity checks. 

Every syscall parameter that is either a pointer, a file or directory name, or an allocated buffer, can be monitored for double fetches. 
As Trinity already knows the data types of all syscall parameters, we can easily extend the main fuzzing routine. 
After Trinity selects a syscall to fuzz, it chooses the arguments to test with and starts a new process. 
Within this process, we spawn a \FlushReload monitoring thread for every parameter that may potentially contain a double-fetch bug. 
The monitoring threads continuously flush the corresponding syscall parameter and measure the reload time. 
As soon as the parameter is accessed from kernel code, the monitoring thread measures a low access time. 
The threads report the number of detected accesses to the referenced memory after the syscall has been executed. 
These findings are logged, and simultaneously, all syscalls with double fetches are added to a candidate set for the interleaved exploitation phase.
In \cref{sec:exploitation}, we additionally show how the second building block \PTwo, allows to automatically test whether such a double fetch is exploitable. 
\Cref{fig:trinity-extended} (\Cref{sec:appendix_dfetches}) shows the process structure of our augmented version of Trinity, called \emph{TrinityDECAF}.

\subsection{Double-Fetch Detection for Black Boxes}\label{sec:properties}

The \FlushReload-based detection method (\POne) is not limited to double fetches in operating system kernels. 
In general, we can apply the technique for all black boxes fulfilling the following criteria:
\begin{compactenum}
 \item Memory references can be passed to the black box.
 \item The referenced memory is (temporarily) shared between the black box and the host.
 \item It is possible to run code in parallel to the execution of the black box. 
\end{compactenum}

This generalization does not only apply to syscalls, but it also applies to trusted execution environments.

Trusted execution environments are particularly interesting targets for double fetch detection and double-fetch-bug exploitation. 
Trusted execution environments isolate programs from other user programs and the operating system kernel.
These programs are often neither open source nor is the unencrypted binary available to the user. 
Thus, if the vendor did not test for double-fetch bugs, researchers not affiliated with the vendor have no possibility to scan for these vulnerabilities. 
Moreover, even the vendor might not be able to apply traditional double-fetch detection techniques, such as dynamic program analysis, if these tools are not available within the trusted execution environment. 

Both Intel SGX~\cite{McKeen2013} and ARM TrustZone~\cite{Alves2004} commonly share memory buffers between the host application and the trustlet running inside the trusted execution environment through their interfaces.
Therefore, we can again utilize a \FlushReload monitoring thread to detect double fetches by the trusted application (\POne). 

\section{Exploiting Double-Fetch Bugs}\label{sec:exploitation}

In this section, we detail the second building block of DECAF, primitive \PTwo, the base of the DECAF exploitation phase.
It allows us to exploit any double fetch found via \POne (\cf \Cref{sec:discovery}) reliably and automatically.
In contrast to state-of-the-art value flipping~\cite{Jurczyk2013} (success probability \SI{50}{\percent} or significantly lower), our exploitation phase has a success probability of \SI{97}{\percent}.
The success probability of value flipping is almost zero if multiple sanity checks are performed, whereas the success probability of \PTwo decreases only slightly.

\subsection{\FlushReload as a Trigger Signal}\label{sec:triggersignal}
We propose to use \FlushReload as a trigger signal to deterministically and reliably exploit double-fetch bugs.
Indeed, \FlushReload is a reliable approach to detect access to the memory, allowing us to flip the value immediately after an access.
This combination of a trigger signal and targeted value flipping forms primitive \PTwo.

The idea of the double-fetch-bug exploitation (\PTwo) is therefore similar to the double-fetch detection (\POne). 
As soon as one access to a parameter is detected, the value of the parameter is flipped to an invalid value. 
Just as the double-fetch detection (\cf \cref{sec:discovery}), we can use a double-fetch trigger signal for every black box which uses memory references as parameters in the communication interface. 

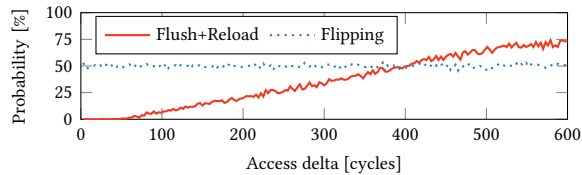
\begin{figure}[t]
 \centering
 \begin{tikzpicture}
\begin{axis}[
style={font=\footnotesize},
xlabel={Access delta [cycles]},
ylabel={Probability [\%]},
ylabel style={text width=2cm,align=center},
width=0.95\hsize,
scaled y ticks=false,
xmin=0,
xmax=600,
ymax=100,
ymin=0,
legend columns=2,
legend pos=north west,
ytick={0,25,50,75,100},
height=3cm
]
\addplot+[red,thick,no marks] table[x=cycles,y=flush,col sep=space] {data/flip_flush.csv};
\addplot+[blue,thick,no marks,dotted] table[x=cycles,y=flip,col sep=space] {data/flip_flush.csv};
\addlegendentry{\FlushReload~\quad~}
\addlegendentry{Flipping}
\end{axis}
\end{tikzpicture}
 \caption{The probability of successfully exploiting a double-fetch bug depending on the time between the accesses.}
 \label{fig:prob-exploit}
\end{figure}

As shown in \Cref{fig:prob-exploit}, the exploitation phase can target double-fetch bugs with an even lower time delta between the accesses, than the double-fetch detection in the profiling phase (\cf \Cref{sec:discovery}).
The reason is that only the first access has to be detected and changing the value is significantly faster than a full \FlushReload cycle.
Thus, it is even possible to exploit double fetches where the time between them is already too short to detect them.
Consequently, every double fetch detected in the profiling phase can clearly be tested for exploitability using \PTwo in the exploitation phase.

As a fast alternative to \FlushReload, \FlushFlush~\cite{Gruss2016Flush} could be used.
\FlushFlush is significantly faster than \FlushReload, but also noisier. 
Our experiments confirmed that \FlushReload is the better choice for reliable double-fetch-bug exploitation.

\subsection{Automated Syscall Exploitation}\label{sec:sub_ase}

With the primitive \PTwo from \Cref{sec:triggersignal}, we add the second building block to DECAF, to not only detect double fetches but also to immediately exploit them. 
This has the advantage that exploitable double-fetch bugs can be found without human interaction, as the automated exploitation leads to evident errors and system crashes. 
As described in \Cref{sec:discovery}, DECAF does not only report the double fetches but also adds them to a candidate set for double-fetch bug testing.
If a candidate is added to this set, the double-fetch bug test (\PTwo) is immediately interleaved into the regular fuzzing process.

We randomly switch between four different methods to change the value: setting it to zero, flipping the least significant bit, incrementing the value, and replacing it by a random value. 
Setting a value to zero or a random value is useful to change pointers to invalid locations. 
Furthermore, it is also effective on string buffers as it can shorten the string, extend the string, or introduce invalid characters. 
Incrementing a value or flipping the least significant bit is especially useful if the referenced memory contains integers, as it might trigger off-by-one errors. 

In summary, in the exploitation phase of DECAF, we reduce the double-fetch candidate set (obtained via \POne) to exploitable double-fetch bugs without any human interaction (\PTwo), complementing state-of-the-art techniques~\cite{Jurczyk2013,Wang2017}.
The coverage of DECAF highly depends on the fuzzer used. 
Fuzzing is probabilistic and might not find every exploitable double fetch, but with growing coverage of fuzzers, the coverage of DECAF will automatically grow as well.

\section{Eliminating Double-Fetch Bugs}\label{sec:prevention}
In this section, we propose the first transparent and automated technique to entirely eliminate double-fetch bugs (\PThree). 
We utilize previously unknown interactions between double-fetch bugs and hardware transactional memory.
\PThree protects code without requiring to identify the actual cause of a double-fetch bug and can even be applied as a preventative measure to protect critical code.

We present the \DropIt library, an instantiation of \PThree with Intel TSX.
\DropIt eliminates double-fetch bugs, having the same effect as rewriting the code to eliminate the double fetch. 
We also show its application to Intel SGX, a trusted execution environment that is particularly interesting in cloud scenarios.

\subsection{Problems of State-of-the-Art Double-Fetch Elimination}\label{sec:problemsdfel}
Introducing double-fetch bugs in software happens easily, and they often stay undetected for many years. 
As shown recently, modern operating systems still contain a vast number of double fetches, some of which are exploitable double-fetch bugs~\cite{Jurczyk2013,Wang2017}.
As shown in \Cref{sec:discovery} and \Cref{sec:exploitation}, identifying double-fetch bugs requires full code coverage, and before our work, a manual inspection of the detected double fetches. 
Even when double-fetch bugs are identified, they are usually not trivial to fix.

A simple example of a double-fetch bug is a syscall with a string argument of arbitrary length.
The kernel requires two accesses to copy the string, first to retrieve the length of the string and allocate enough memory, and second, to copy the string to the kernel. 

Writing this in a na\"ive way can lead to severe problems, such as unterminated strings of kernel buffer overflows. 
One approach is to use a retry logic, as shown in \Cref{alg:string-cpy} (\Cref{sec:appendix_strcpy}), as it used in the Linux kernel whenever user data of unknown length has to be copied to the kernel. 
Such methods increase the complexity and runtime of code, and they are hard to wrap into generic functions. 

Finally, compilers can also introduce double fetches that are neither visible in the source code nor easily detectable, as they are usually within just a few cycles~\cite{Blanchou2013}.

\subsection{Generic Double-Fetch Bug Elimination}\label{sec:gdfbe}
Eliminating double-fetch bugs is not equivalent to eliminating double fetches. 
Double fetches are valid constructs, as long as a change of the value is successfully detected, or it is not possible to change the value between two memory accesses. 
Thus, making a series of multiple fetches atomic is sufficient to eliminate double-fetch bugs, as there is only one operation from an attacker's view (see \cref{sec:tsx}). 
Curiously, the concept of hardware transactional memory provides exactly this atomicity.

As also described in \Cref{sec:tsx}, transactional memory provides atomicity, consistency, and isolation~\cite{Harris2010}. 
Hence, by wrapping code possibly containing a double fetch within a hardware transaction, we can benefit from these properties. 
From the view of a different thread, the code is one atomic memory operation. 
If an attacker changes the referenced memory while the transaction is active, the transaction aborts and can be retried.
As the retry logic is implemented in hardware and not simulated by software, the induced overhead is minimal, and the amount of code is drastically reduced.

In a nutshell, hardware transactional memory can be instrumented as a hardware implementation of software-based retry solutions discussed in \Cref{sec:problemsdfel}.
Thus, wrapping a double-fetch bug in a hardware transaction does not hide, but actually eliminates the bug (\PThree).
Similar to the software-based solution, our generic double-fetch bug elimination can be automatically applied in many scenarios, such as the interface between trust domains (\eg \texttt{ECALL} in SGX). 
Naturally, solving a problem with hardware support is more efficient, and less error-prone, than a pure software solution.

In contrast to software-based retry solutions, our hardware-assisted solution (\PThree) does not require any identification of the resource to be protected.
For this reason, we can even prevent undetectable or yet undetected double-fetch bugs, regardless of whether they are introduced on the source level or by the compiler.
As these interfaces are clearly defined, the double-fetch bug elimination can be applied in a transparent and fully automated manner. 

\subsection{Implementation of \DropIt} 
To build \DropIt, our instantiation of \PThree, we had to rely on real-world hardware transactional memory, namely Intel TSX.
Intel TSX comes with a series of imperfections, inevitably introducing practical limitations for security mechanisms, as observed in previous work~\cite{Gruss2017Cloak} (\cf \Cref{sec:tsx}).
However, as hardware transactional memory is exactly purposed to make multiple fetches from memory consistent, Intel TSX is sufficient for most real-world scenarios.

To eliminate double-fetch bugs, \DropIt relies on the \texttt{XBEGIN} and \texttt{XEND} instructions of Intel TSX. 
\texttt{XBEGIN} specifies the start of a transaction as well as a fall-back path that is executed if the transaction aborts, whereas \texttt{XEND} marks the successful completion of a transaction. 

We find that on a typical Ubuntu Linux the kernel usually occupies less than 32\,MB including all code, data, and heap used by the kernel and kernel modules.
With an 8\,MB L3 cache we could thus read or execute more than 20\,\% of the kernel without causing high abort rates~\cite{Gruss2017Cloak} (\cf Section~\ref{sec:tsx}).
In Section~\ref{sec:eval_dropit}, we show that for practical use cases the abort rates are almost 0\,\% and our approach even improves the system call performance in several cases.

\DropIt abstracts the transactional memory as well as the retry logic from the programmer.
Hence, in contrast to existing software-based retry logic (\cf \Cref{sec:problemsdfel}), \eg in the Linux kernel, \DropIt is mostly transparent to the programmer.
To protect code, \DropIt takes the number of automatic retries as a parameter as well as a fall-back function for the case that the transaction is never successful, \ie for the case of an ongoing attack. 
Hence, a programmer only has to add 3 lines of code to protect arbitrary code from double fetch exploitation.
Listing~\ref{lst:drop-it} (\Cref{sec:appendix_dropit}) shows an example how to protect the insecure \texttt{strcpy} function using \DropIt. 
The solution with DropIt is clearly simpler than current state-of-the-art software-based retry logic (\cf \Cref{alg:string-cpy}). 
Finally, replacing software-based retry logic by our hardware-assisted \DropIt library can also improve the execution time of protected syscalls.

\DropIt is implemented in standard C and does not have any dependencies. 
It can be used in user space, kernel space, and in trusted environments such as Intel SGX enclaves. 
If TSX is not available, \DropIt immediately executes the fall-back function.
This ensures that syscalls still work on older systems, while modern systems additionally benefit from possibly increased performance and elimination of yet unknown double-fetch bugs. 

\DropIt can be used for any code containing multiple fetches, regardless of whether they have been introduced on a source-code level or by the compiler. 
In case there is a particularly critical section in which a double fetch can cause harm, we can automatically protect it using \DropIt.
For example, this is possible for parts of syscalls that interact with the user space. 
As these parts are known to a compiler, a compiler can simply add the \DropIt functions there. 

\DropIt is able to eliminate double-fetch bugs in most real-world scenarios. 
As Intel TSX is not an ideal implementation of hardware transactional memory, use of certain instructions in transactions is restricted, such as port I/O instructions~\cite{Intel_ISA}.
However, double fetches are typically caused by string handling functions and do not rely on any restricted instructions. 
Especially in a trusted environment, such as Intel SGX enclaves, where I/O operations are not supported, all functions interacting with the host application can be automatically protected using \DropIt. 
This is a particularly useful protection against an attacker in a cloud scenario, where an enclave containing an unknown double-fetch bug may be exposed to an attacker over an extended period of time.

\section{Evaluation}\label{sec:evaluation}

The evaluation consists of four parts. The first part evaluates DECAF (\POne and \PTwo), the second part compares \PTwo to state-of-the-art exploitation techniques, the third part evaluates \POne on trusted execution environments, and the fourth part evaluates \DropIt (\PThree).

First, we demonstrate the proposed detection method using \FlushReload. 
We evaluate the double-fetch detection of TrinityDECAF on both a recent Linux kernel 4.10 and an older Linux kernel 4.6 on Ubuntu 16.10 and discuss the results. 
We also evaluate the reliability of using \FlushReload as a trigger in TrinityDECAF to exploit double-fetch bugs (\PTwo). 
On Linux 4.6, we show that TrinityDECAF successfully finds and exploits CVE-2016-6516, a real-world double-fetch bug.

Second, we compare our double-fetch bug exploitation technique (\PTwo) to state-of-the-art exploitation techniques.
We show that \PTwo outperforms value-flipping as well as a highly optimized exploit crafted explicitly for one double-fetch bug.
This underlines that \PTwo is both generic and extends the state of the art significantly.

Third, we evaluate the double-fetch detection (\POne) on trusted execution environments, \ie Intel SGX and ARM TrustZone.
We show that despite the isolation of those environments, we can still use our techniques to detect double fetches.

Fourth, we demonstrate the effectiveness of \DropIt, our double-fetch bug elimination method (\PThree). 
We show that \DropIt eliminates source-code-level double-fetch bugs with a very low overhead. 
Furthermore, we reproduce CVE-2015-8550, a compiler-introduced double-fetch bug.
Based on this example we demonstrate that \DropIt also eliminates double-fetch bugs which are not even visible in the source code. 
Finally, we measure the performance of \DropIt protecting 26 syscalls in the Linux kernel, where TrinityDECAF reported double fetches.

\subsection{Evaluation of DECAF}\label{sec:eval-decaf}
To evaluate DECAF, we analyze the double fetches and double-fetch bugs reported by TrinityDECAF.
Our goal here is not to fuzz an excessive amount of time, but to demonstrate that DECAF constitutes a sensible and practical complement to existing techniques.
Hence, we also used old and stable kernels where we did not expect to find new bugs, but validate our approach.

\paragraph{Reported Double-Fetch Bugs}
Besides many double fetches TrinityDECAF reports in Linux kernel 4.6, it identifies one double-fetch bug which is already documented as CVE-2016-6516. 
It is a double-fetch bug in one of the \texttt{ioctl} calls.
The syscall is used to share physical sections of two files if the content is identical. 

When calling the syscall, the user provides a file descriptor for the source file as well as a starting offset and length within the source file.
Furthermore, the syscall takes an arbitrary number of destination file descriptors with corresponding offsets and lengths. 
The kernel checks whether the given destination sections are identical to the source section and if this is the case, frees the sections and maps the source section into the destination file. 

As the function allows for an arbitrary number of destination files, the user has to supply the number of provided destination files. 
This number is used to determine the amount of memory required to allocate. 
Listing~\ref{lst:filededuperange} (\Cref{sec:appendix_vuln}) shows the corresponding code from the Linux kernel. 
Changing the number between the allocation and the actual access to the data structure leads to a kernel heap-buffer overflow. 
Such an overflow can lead to a crash of the kernel or even worse to a privilege escalation. 

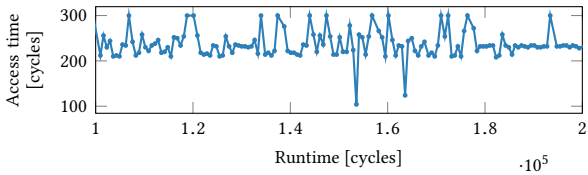
\begin{figure}[t]
 \centering
\begin{tikzpicture}
\begin{axis}[
mlineplot,
style={font=\footnotesize},
xlabel={Runtime [cycles]},
ylabel={Access time [cycles]},
ylabel style={text width=2cm,align=center},
width=0.95\hsize,
scaled y ticks=false,
xmin=100000,
xmax=200000,
height=3cm
]
\addplot+[blue,thick,mark options={draw=blue,fill=blue},mark=*,mark size=0.5] table[x=Time,y=Access,col sep=space] {data/double_fetch.csv};

\end{axis}
\end{tikzpicture}
 \caption{A real-world double-fetch bug that was present in the Linux kernel from version 4.5 to 4.7. 
 The two memory accesses of the vulnerable function in the \texttt{FIDEDUPERANGE} \texttt{ioctl} can be clearly seen at around $1.5 \cdot 10^5$ and $1.6 \cdot 10^5$ cycles.}
 \label{fig:double-fetch-ioctl}
\end{figure}

Trinity already has rudimentary support for the \texttt{ioctl} syscall, which we extended with semi-intelligent defaults for parameter selection.
Consequently, while Trinity does not find CVE-2016-6516, TrinityDECAF indeed successfully detects this double fetch in the profiling phase.
\cref{fig:double-fetch-ioctl} shows a cache trace while calling the vulnerable function on Ubuntu 16.04 running an affected kernel 4.6. 
Although the time between the two accesses is only \SIx{10000} cycles (approximately \SI{2.5}{\micro\second} on our i7-6700K test machine), we can clearly detect the two memory accesses. 

When, in the exploitation phase, the monitoring thread changes the value to a higher value (\cf \Cref{sec:sub_ase}) exceeding the actual number of provided file descriptors, the kernel iterates out-of-bounds, as the number of file descriptors does not match the actual number of file descriptors anymore. 
This out-of-bounds access to the heap buffer results in a denial-of-service of the kernel and thus a hard reboot is required. 
Consequently, the denial-of-service shows that the double fetch is an exploitable double-fetch bug.

This demonstrates that DECAF is a useful complement to state-of-the-art fuzzing techniques, allowing to automatically detect bugs that cannot be found with traditional fuzzing approaches.

\paragraph{Reported Double Fetches}
Besides Linux kernel 4.6, we also tested TrinityDECAF on a recent Linux kernel 4.10.
We let TrinityDECAF investigate all 64-bit syscalls (currently \SIx{295}) without exceptions for one hour on an Intel i7-6700K. 
On average, every syscall was executed \SIx{8058} times. 
Due to the semi-intelligent parameter selection of TrinityDECAF, most syscalls are called with valid parameters. 
In our test run, \SI{75.12}{\percent} of the syscalls executed successfully. 
Hence, on average, every syscall was successfully executed \SIx{6053} times, indicating a high code coverage for every syscall. 

For every syscall parameter, TrinityDECAF displays a percentage of the calls where it detected a double fetch. 
Out of the \SIx{295} tested 64-bit syscalls, TrinityDECAF reported double fetches for \SIx{68} syscalls in Linux kernel 4.10. 
This is not surprising and in line with state-of-the-art work~\cite{Wang2017} which reported \SIx{90} double fetches in Linux, but only \SIx{33} in syscalls.
For each of the reported syscalls, we investigated the respective implementation. 
\cref{tab:syscall-dfetches} (\Cref{sec:appendix_dfetches}) shows a complete list of reported syscalls and the reason why TrinityDECAF detected a double fetch. 
We can group the reported syscalls into 5 major categories, explaining the detected double fetch.

\begin{itemize}[nolistsep,align=left, leftmargin=9pt, labelwidth=0pt, itemindent=!]
  \item \textbf{Filenames.} 
  Most syscalls handling filenames (or paths) are reported by TrinityDECAF. 
  Many of them use \texttt{getname\_flags} internally to copy a filename to a kernel buffer. 
  This function checks whether the filename is already cached in the kernel, and copies it to the kernel if this is not the case, resulting in multiple accesses to the file name. 
  The exploitation phase automatically filtered out all non-exploitable double fetches in this category.
  
 \item \textbf{Shared input/output parameters.} 
 We found 5 syscalls which are reported by TrinityDECAF although they do not contain a double fetch. 
 In these syscalls, one of the syscall parameters was used as input and output. 
 As reads and writes are not distinguishable through the cache access pattern (\cf \cref{sec:classification}), these syscalls are filtered out automatically in the exploitation phase.
 
 \item \textbf{Strings of arbitrary length.}
 As with filenames, some syscalls expect strings from the user that do not have a fixed length. 
 To safely copy such arbitrary length strings, some syscalls (\eg \texttt{mount}) use an algorithm similar to \cref{alg:string-cpy}. 
 Thus, the detected double fetch is due to the length check and the subsequent string copy. 
  The exploitation phase automatically filtered out all non-exploitable double fetches in this category.
 
 \item \textbf{Sanity check.}
 Many syscalls check---either directly, or in a sub-routine---whether the supplied argument is sane. 
 There are sanity checks that check whether it is safe to access a user-space pointer before actually copying data from or to it. 
 Such a check can also trigger a cache hit if the value was actually accessed. 
 All correct sanity checks were automatically filtered out in the exploitation phase.
 The exploitation phase correctly identified the \texttt{ioctl} syscall in the Linux kernel 4.6, but also correctly filtered it out in Linux kernel 4.10.
 
 \item \textbf{Structure elements.} 
 If a syscall has a structure as parameter, double fetches can be falsely detected if structure members fall into the same cache line (\cf \cref{sec:classification}). 
 If members are either copied element-wise or neighboring members are simply accessed, TrinityDECAF will detect a double fetch although two different variables are accessed. 
 Again, these false positives are filtered out in the exploitation phase.
\end{itemize}

Our evaluation showed that TrinityDECAF provides a sensible complement to existing double-fetch bug detection techniques.
The fact that we found only \SIx{1} exploitable double-fetch bug in \SIx{68} double fetches is not surprising, and in line with previous work, \eg Wang~\etal\cite{Wang2017} found \SIx{3} exploitable double-fetch bugs by manually inspecting \SIx{90} double fetches they found.
However, it also shows that the coverage of DECAF highly depends on the fuzzer used to instantiate it.
Future work may retrofit other fuzzers with DECAF, to extend the spectrum of bugs that the fuzzer covers and thereby also extend the coverage of DECAF.
Furthermore, as Trinity is continuously extended, the coverage of TrinityDECAF grows automatically with the coverage of Trinity.

\subsection{Evaluation of \PTwo}

\begin{figure}[t]
 \centering
 \begin{tikzpicture}
  \begin{axis}[
      ybar,       
      height=3.2cm,
      width=\hsize,
      ymax=125,
      ymin=-5,
	 enlarge x limits={abs=1cm},
      legend style={at={(0.5,-0.3)},
	  anchor=north,legend columns=-1,draw=none},
	  ytick={0,50,100},
      yticklabels={0\%,50\%,100\%},
      symbolic x coords={Flipping,Wait,Flush+Reload},
      xtick=data,
      nodes near coords,
      nodes near coords align={vertical},
      ]
  \addplot [fill=red] coordinates {(Flipping,49) (Wait,12) (Flush+Reload,0)};
  \addplot [fill=green] coordinates {(Flipping,25) (Wait,84) (Flush+Reload,97)};
  \addplot [fill=yellow!50!white] coordinates {(Flipping,26) (Wait,4) (Flush+Reload,3)};
  \legend{too early \qquad,success \qquad,too late}
  \end{axis}
\end{tikzpicture}
 \caption{Comparing three exploits for the double-fetch bug CVE-2016-6516. 
 Our \FlushReload-based trigger in TrinityDECAF succeeds in \SI{97}{\percent} of the cases, outperforming the provided proof-of-concept (\SI{84}{\percent}) and the state-of-the-art method of value flipping (\SI{25}{\percent}).}
 \label{fig:exploits}
\end{figure}
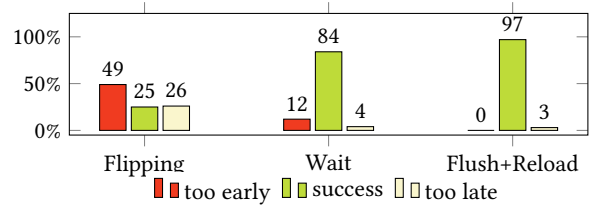

To evaluate \PTwo in detail, we compare three different variants to exploit the double-fetch bug reported in CVE-2016-6516. 
First, the provided exploit, which calls \texttt{ioctl} multiple times, always changing the affected variable after a slightly increased delay. 
Second, we use state-of-the-art value flipping to switch the affected variable as fast as possible between the valid and an invalid value. 
Third, the automated approach \PTwo, integrated into TrinityDECAF.

\cref{fig:exploits} shows the success rate of \SIx{1000} executions of each of the three variants. 
Value flipping has by far the worst success rate, although in theory, it should have a success rate of approximately \SI{50}{\percent}. 
In half of the cases, the value is flipped before the first access. Thus, the exploit fails, as the value is smaller at the next access. 
In the other cases, the probability to switch the value at the correct time is again \SI{50}{\percent} resulting in an overall success rate of \SI{25}{\percent}.

The original exploit is highly optimized for this specific vulnerability. 
It uses a trial-and-error busy wait with steadily increasing timeouts, which works surprisingly well, as there is sufficient time between the two accesses. 
Depending on the scheduling, the attacker sometimes sleeps too long (\SI{4}{\percent}) and sometimes too short (\SI{12}{\percent}). 
Still, the busy wait outperforms the value flipping in this scenario, increasing the success probability from \SI{25}{\percent} to \SI{84}{\percent}. 

Even though our \FlushReload-based trigger (\PTwo) is generic and does not require fine-tuning of the sleep intervals, it has the highest success rate. 
There is no case where the value was changed too early, as there are no false positives with \FlushReload in this scenario. 
Furthermore, as the time between the two memory accesses is long enough, we achieve an almost perfect success rate of \SI{97}{\percent}. 
The remaining \SI{3}{\percent}, where we do not trigger a change of the value, are caused by unfortunate scheduling of the application. 

\begin{figure}[t]
 \centering
 \resizebox{\hsize}{!}{
 \begin{tikzpicture}
\begin{axis}[
    legend pos=outer north east,
    mlineplot,
    xlabel={Number of checks},
    xtick={1,2,3,4},
    xticklabels={1,2,3,4},
    ytick={0,25,50,75,100},
    yticklabels={0,25,50,75,100},
    ylabel={Probability [\%]},
	ylabel style={align=center,text width=1cm},
    width=0.75\hsize,
    scaled y ticks=false,
    xmin=1,
    xmax=4,
    ymin=0,
    ymax=100,
    ymajorgrids,
    xmajorgrids,
    height=3cm
]
\addplot+[red,thick,no marks] table[x=fetches,y=cache,col sep=space] {data/multi.csv};
\addplot+[dashed,blue,thick,no marks] table[x=fetches,y=flipping,col sep=space] {data/multi.csv};
\addlegendentry{\FlushReload}
\addlegendentry{Flipping}
\end{axis}
\end{tikzpicture}
 }
 \caption{The probability to exploit a double-fetch bug if a certain number of fetches are used for sanity check.
 The more sanity checks, the lower the probability to change the value only between the second to last and last access.}
 \label{fig:multi-fetch}
\end{figure}
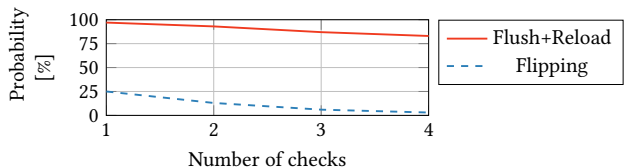

The success rate of value flipping drops significantly if the two values have to fulfill specific constraints, \eg the value has to be higher on the second access. 
For example, if an application does not only fetch the value twice, but multiple times for sanity checking, the probability of successfully exploiting it using value flipping decreases exponentially. 

\cref{fig:multi-fetch} shows the probability to exploit a double fetch similar to CVE-2016-6516, with additional fetches for sanity checks. 
To successfully exploit the vulnerability, the value has to be the same for all sanity checks and must be higher for the last access. 
Flipping the value between a valid and an invalid value decreases the chances by \SI{50}{\percent} for every additional sanity check. 

Our \FlushReload-based method (\PTwo) does not suffer significantly from additional sanity checks. 
We can accurately trigger on the second to last access to change the value. 
The slightly decreased probability is only due to missed accesses.

\subsection{Evaluation of \POne on Trusted Execution Environments}\label{sec:eval-tees}
We evaluate \POne on trusted execution environments by successfully detecting double fetches in Intel SGX and ARM TrustZone.

\paragraph{Intel SGX}

Intel SGX allows running code in secure enclaves without having to trust the user or the operating system. 
A program running inside an enclave is not accessible to the operating system due to hardware isolation provided by SGX. 
Weichbrodt~\etal\cite{Weichbrodt2016} showed that synchronization bugs, such as double fetches, inside SGX enclaves, can be exploited to hijack the control flow or bypass access control. 

As it has been shown recently, enclaves leak information through the last-level cache, even to unprivileged user space applications, as they share the last-level cache with regular user space applications~\cite{Gotzfried2017,Schwarz2017SGX,Brasser2017sgx,Moghimi2017}.
SGX enclaves provide a communication interface using so-called \texttt{ecall}s and \texttt{ocall}s, similar to the syscall interface. 
Enclaves fulfill the properties of \cref{sec:properties}, and we can thus detect double fetches within enclaves, even without access to the binary. 
Therefore, we can apply our method to identify double fetches within SGX enclaves.

To test our \FlushReload detection mechanism (\POne), we implemented a small enclave application. 
This application consists of only one \texttt{ecall}, which takes a memory reference as a parameter. 
As enclaves can access non-enclave memory, the user can simply allocate memory and provide the pointer to the enclave. 
The enclave accesses the memory once, idles a few thousand cycles and reaccesses the memory. 
Although the enclave should be isolated from other applications, the monitoring application can clearly detect the 2 cache hits. 
\cref{fig:double-fetch-sgx} (\Cref{sec:appendix_trustzone}) shows the measurement of the \FlushReload thread running outside the enclave on an Intel i5-6200U. 
Similarly, \Cref{sec:appendix_trustzone} evaluates \POne on ARM TrustZone. 

\subsection{Evaluation of \DropIt}\label{sec:eval_dropit}
To evaluate our open-source library \DropIt, as an instantiation of \PThree, we investigate two real-world scenarios.
In the first scenario, we demonstrate how \DropIt eliminates a compiler-introduced real-world double-fetch bug in Xen (CVE-2015-8550). 
In the second scenario, we evaluate the effect of \DropIt on Linux syscalls with double fetches.
Our findings show that \DropIt successfully eliminates all double-fetch bugs and can be used as a preventative measure to protect double fetches in syscalls generically.

\paragraph{Eliminating Compiler-Introduced Double-Fetch Bugs}
As discussed in \Cref{sec:dfadfb}, compilers can also introduce double-fetch bugs.
Especially switch statements are prone to double-fetch bugs if the variable is subject to a race condition~\cite{CWE-365,Blanchou2013}. 
This is not an issue with the compiler, as the compiler is allowed to assume an atomic value for the switch condition~\cite{ANSI1999}.
We are aware of two scenarios where code generated by gcc contains a double-fetch bug. 

If a switch is translated into a jump table with a default case, gcc generates two accesses to the switch variable. 
The first access checks whether the parameter is within the bounds of the jump table, the second access calculates the actual jump target. 
Thus, if the parameter changes between the accesses, a malicious user can divert the control flow of the program. 

If the switch is implemented as multiple conditional jumps, the compiler is allowed to fetch the variable for every conditional jump. 
This leads to cases where the switch executes the default case as the variable changes while checking the conditions~\cite{CWE-365}. 

We evaluated \DropIt on the real-world compiler-introduced double-fetch bug CVE-2015-8550. 
This vulnerability in Xen allowed arbitrary code execution due to a compiler-introduced double fetch within a switch statement.
Note that such a switch statement is a common construct and can occur in any other kernel, \eg Linux, or Windows, if a memory buffer is shared between user space and kernel space.
Wrapping the switch statement using \DropIt results in a clean and straightforward fix without relying on the compiler. 
With \DropIt, any compiler-introduced switch-related double-fetch bug is successfully eliminated using only 3 lines of additional code.

To compare the overhead of traditional locking and \DropIt, we implemented a minimal working example of a compiler-introduced double-fetch bug. 
Our example consists of a switch statement that has 5 different cases as well as a default case. 
The condition is a pointer which is subject to a race condition. 
The average execution time of the switch statement without any protection is \SIx{7.6} cycles. 
Using a spinlock to protect the variable increased the average execution time to \SIx{83.7} cycles. 
\DropIt achieved a higher performance than the traditional spinlock with an average execution time of \SIx{68.0} cycles. 
Thus, with 3 additional lines of code, \DropIt is not only easy to deploy but also achieves a better performance than traditional locking mechanisms. 

\paragraph{Preventative Protection of Linux Syscalls}
To show that \DropIt provides an automated and transparent generic solution to eliminate double-fetch bugs, we also used \DropIt in the Linux kernel. 
As discussed in \cref{sec:eval-decaf}, a majority of the double fetches we detected in the Linux kernel are due to the \texttt{getname\_flags} function handling file names. 
We replaced this function with a straight-forward implementation protected by \DropIt. 
With this small change, all double fetches previously reported in 26 syscalls were covered by \DropIt, and thus all potential double-fetch bugs were eliminated.

\begin{figure}[t]
 \centering
 \resizebox{\hsize}{!}{
 \begin{tikzpicture}
\begin{axis}[
	mbarplot,
    width=12cm,
    height=3.5cm,
    xtick={1,...,15},
    xticklabels={%
mkdir,
chdir,
rmdir,
create,
open,
read,
write,
close,
stat,
access,
chmod,
readdir,
link,
unlink,
delete
},
     xticklabel style={rotate=70},
    ybar,
	ylabel = {Operations/s}
    ]

\addplot[
    fill=green,
    draw=black,
    point meta=y,
    every node near coord/.style={inner ysep=5pt},
    error bars/.cd,
        y dir=both,
        y explicit
] 
table [y=TSX,y error=errTSX,col sep=space] {data/fileio.csv};

\addplot[
    fill=red,
    draw=black,
    point meta=y,
    every node near coord/.style={inner ysep=5pt},
    error bars/.cd,
        y dir=both,
        y explicit
] 
table [y=noTSX,y error=errnoTSX,col sep=space] {data/fileio.csv};

\draw ({rel axis cs:0,0}|-{axis cs:0,0}) -- ({rel axis cs:1,0}|-{axis cs:0,0});
\end{axis}
\end{tikzpicture}
 }
 \caption{IOZone Fileops benchmark showing the performance of file operations. 
 The number of executed operations per second of our re-implemented \texttt{getname\_flags} using \DropIt (green) does not significantly differ from the version in the vanilla kernel (red).}
 \label{fig:fileio-benchmark} 
\end{figure}
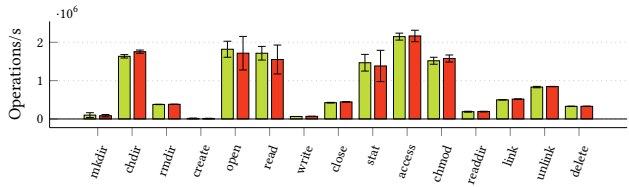

To compare the performance of \DropIt with the vanilla implementation, we executed \SIx{210} million file operations in both cases. 
All benchmarks were run on a bare metal kernel to reduce the impact of system noise. 
\cref{fig:fileio-benchmark} shows the result of the IOzone filesystem benchmark~\cite{Norcott2016}. 
On average, the benchmarks show a \SI{\ValDropItOverhead}{\percent} performance degradation on the tested file operations that are affected by our kernel change. 
In some cases, \DropIt even has a better performance than the vanilla implementation. 
We therefore conclude that \DropIt has no perceptible performance impact. 
The variances in the tests are probably due to the underlying hardware, \ie the SSDs on which we performed the file operations. 

Thus, \DropIt provides a reliable and straightforward way to cope with double-fetch bugs. 
It is easily integrable into existing C projects and does not negatively influence the performance compared to state-of-the-art solutions. 
Furthermore, it even increases the performance compared to traditional locking mechanisms.

\section{Conclusion}\label{sec:conclusion}

In this paper, we proposed novel techniques to efficiently detect, exploit, and eliminate double-fetch bugs.
We presented the first combination of state-of-the-art cache attacks with kernel-fuzzing techniques.
This allowed us to find double fetches in a fully automated way.
Furthermore, we presented the first fully automated reliable detection and exploitation of double-fetch bugs.
By combining these two primitives, we built DECAF, a system to automatically find and exploit double-fetch bugs.
DECAF is the first method that makes manual analysis of double fetches as in previous work superfluous.
We show that cache-based triggers, as we use in DECAF, outperform state-of-the-art exploitation techniques significantly, leading to an exploitation success rate of up to 97\,\%. 

DECAF constitutes a sensible complement to existing double-fetch detection techniques.
Future work may retrofit more fuzzers with DECAF, extending the spectrum of bugs covered by fuzzers.
Hence, double-fetch bugs do not require separate detection tools anymore, but testing for these bugs can now be a part of regular fuzzing.
With continuously growing coverage of fuzzers, the covered search space for potential double-fetch bugs grows as well.

With \DropIt, we leverage a newfound interaction between hardware transactional memory and double fetches, to completely eliminate double-fetch bugs.
Furthermore, we showed that \DropIt can be used in a fully automated manner to harden Intel SGX enclaves such that double-fetch bugs cannot be exploited. 
Finally, our evaluation of \DropIt in the Linux kernel showed that it can be applied to large systems with a negligible performance overhead below \SI{1}{\percent}.


\bibliographystyle{IEEEtranS}
\bibliography{ms}

\FloatBarrier

\appendix

\section{TrinityDECAF and Detected Double Fetches}\label{sec:appendix_dfetches}

\FloatBarrier

In this section, we show implementation details of TrinityDECAF (our augmented version of Trinity) as well as a complete list of syscalls reported by TrinityDECAF. 

\Cref{fig:trinity-extended} shows the process structure of our augmented version of Trinity, called \emph{TrinityDECAF}. 
The syscall fuzzer Trinity is extended with one monitoring threads per syscall argument. 
Each of the monitoring threads mounts a \FlushReload attack to detect double fetches (\cf \Cref{sec:trinitydecaf_syscalls}).

\begin{figure}[!hb]
 \centering
 \begin{tikzpicture}
\begin{axis}[
mlineplot,
style={font=\footnotesize},
xlabel={Access delta [cycles]},
ylabel={Probability [\%]},
width=0.95\hsize,
scaled y ticks=false,
xmin=200,
xmax=3500,
height=3cm
]
\addplot+[blue,thick,no marks] table[x=cycles,y=prob,col sep=space] {data/prob_double_fetch.csv};

\end{axis}
\end{tikzpicture}
 \caption{The probability of detecting a double fetch depending on the time between the accesses.}
 \label{fig:double-fetch}
\end{figure}
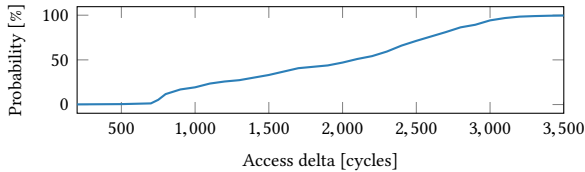

\Cref{fig:double-fetch} shows the probability that TrinityDECAF detects a double fetch depending on the time between the two accesses to the memory (\cf \Cref{sec:prob_df_detection})

\Cref{tab:syscall-dfetches} is a complete table of reported syscalls and the reason why TrinityDECAF detected a double fetch. 
The categories are discussed in detail in \Cref{sec:eval-decaf}.

\begin{figure}[!hb]
 \centering
 \resizebox{0.8\hsize}{!}{
 \tikzstyle{process} = [rectangle, rounded corners, minimum width=2.5cm, minimum height=.6cm,text centered, draw=black, fill=white]
\tikzstyle{processdis} = [rectangle, rounded corners, minimum width=2.5cm, minimum height=.6cm,text centered, draw=black!60!white, fill=white!90!black,text=black!60!white]
\tikzstyle{thread} = [rectangle, minimum width=2cm, minimum height=.6cm, text centered, draw=black, fill=orange!30]
\tikzstyle{arrow} = [thick,->,>=stealth]
\tikzstyle{arrowdis} = [thick,->,>=stealth,draw=black!60!white,fill=black!60!white]

\usetikzlibrary{shapes.geometric, arrows}

\resizebox{0.8\hsize}{!}{
\begin{tikzpicture}[node distance=1.15cm,scale=0.9,transform shape]

\node (trinity) [process] {TrinityDECAF};
\node (trinitymain) [process, below of=trinity, xshift=-7em] {trinity-main};
\node (watchdog) [process, right of=trinitymain, xshift=7em] {trinity-watchdog};
\node (child1) [process, below of=trinitymain, xshift=-7em] {do\_syscall 1};
\node (child2) [processdis, right of=child1, xshift=6em] {do\_syscall 2};
\node (childn) [processdis, right of=child2, xshift=6em] {do\_syscall $n$};
\node (param1) [thread, below of=child1, xshift=1em] {monitor 1};
\node (paramn) [right of=param1, xshift=1em,yshift=0em] {...};
\node (param6) [thread, right of=param1, xshift=5em,yshift=0em] {monitor $m$};

\draw [arrow] (trinity) -- (trinitymain);
\draw [arrow] (trinity) -- (watchdog);
\draw [arrow] (trinitymain) -- (child1);
\draw [arrowdis] (trinitymain) -- (child2);
\draw [arrowdis] (trinitymain) -- (childn);

\draw [arrow, bend left=15] (child1.south) to node[below,sloped] {} (param1.north);
\draw [arrow, bend left=5] (child1.south) to node[below,sloped] {} (param6.north);

\end{tikzpicture}
}
 }
 \caption{The structure of TrinityDECAF with the \FlushReload monitoring threads for the syscall parameters.}
 \label{fig:trinity-extended}
\end{figure}
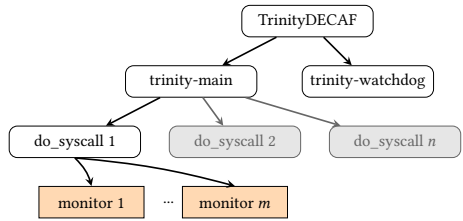

\begin{table}[!hb]
  \centering
 \caption{Double fetches found by TrinityDECAF.}
 \label{tab:syscall-dfetches}
	\small
 \begin{tabular}{lp{5.3cm}}
  \toprule
  Category & Syscall \\
  \midrule
  Filenames & open, newstat, truncate, chdir, rename(at), mkdir(at), rmdir, creat, unlink, link, symlink(at), readlink(at), chmod, (l)chown, utime, mknod, statfs, chroot, quotactl, *xattr, fchmodat \\
  \midrule
  Shared input/output & sendfile, adjtimex, io\_setup, recvmmsg, sendmmsg \\
  \midrule
  Strings & mount, memfd\_create \\
  \midrule
  Sanity check & sched\_setparam, ioctl, sched\_setaffinity, io\_cancel, sched\_setscheduler, futimesat, sysctl, settimeofday, gettimeofday \\
  \midrule
  Structure elements & recvmsg, msgsnd, sigaltstack, utime \\
  \bottomrule
 \end{tabular}
\end{table}

\FloatBarrier

\section{Safe String Copy}\label{sec:appendix_strcpy}

In this section, we show the pseudo-code of a standard algorithm used to safely copy an arbitrary-length string. 
\Cref{alg:string-cpy} first retrieves the length of the string, to allocate a buffer and copy the string up to this length. 
Then, it checks whether the string is terminated, and if not, retries again as the buffer was apparently changed before copying it. 

A similar algorithm is used in the Linux kernel whenever user data of unknown length has to be copied to the kernel. 

\begin{algorithm}
	\SetKw{KwStep}{step}
	\SetKwInOut{Input}{input}\SetKwInOut{Output}{output}
	\Input{$\mathit{string}$}
	copy: \\
	len $\gets$ strlen(string)\;
	buffer $\gets$ allocate(len + 1)\;
	strncpy(buffer, string, len)\;
	\If{$\textbf{not}$ isNullTerminated(buffer)} {
	  free(buffer)\;
	  goto copy; // or abort with error if too many retries
	} 
	\caption{Safe string copy for arbitrary string lengths with software-based retry logic.}
  \label{alg:string-cpy}
\end{algorithm}

\FloatBarrier

\section{CVE-2016-6516}\label{sec:appendix_vuln}

CVE-2016-6516 is a double-fetch bug in an \texttt{ioctl} call used to share physical sections of two files if the content is identical. 
This deduplicates the identical section to save physical storage. 
On a write access, the identical section has to be copied to ensure that the changes are only visible within the changed file. 

The user provides a file descriptor for the source file as well as a starting offset and length within the source file.
Additionally, the syscall takes an arbitrary number of destination file descriptors including offsets and lengths. 
The kernel maps the source section into the destination file if the given destination sections are identical to the source section.

\begin{lstlisting}[float,language=C,caption={The vulnerable \texttt{ioctl}\texttt{\_}\texttt{file}\texttt{\_} \texttt{dedupe}\texttt{\_}\texttt{range} function that was present in the Linux kernel from version 4.5 to 4.7.
The \texttt{dest\_count} member is accessed twice and can thus be changed between the accesses by a malicious user, leading to a kernel heap-buffer overflow.},label={lst:filededuperange}]
// first access of dest_count 
if (get_user((*@\hl{\texttt{count}}@*), (*@ \hl{\texttt{\&argp->dest\_count}}@*))) { [...] }
// allocation based on dest_count
size = offsetof(struct file_dedupe_range __user, 
  info[(*@\hl{\texttt{count}}@*)]);
same = memdup_user(argp, size);
if (IS_ERR(same)) { [...] }
ret = vfs_dedupe_file_range(file, (*@\hl{\texttt{same}}@*)); 
// function accesses (*@\hl{\texttt{same->dest\_count}}@*), not (*@\hl{\texttt{count}}@*)
\end{lstlisting}

The function supports an arbitrary number of destination files. 
Thus, the user has to supply the number of provided destination files, so that the kernel can determine the required amount of memory to allocate. 
Listing~\ref{lst:filededuperange} shows the corresponding code from the Linux kernel. 
If the number changes between the allocation and the actual access to the data structure, the kernel accesses the buffer out-of-bounds, leading to a heap-buffer overflow.

\FloatBarrier

\section{ARM TrustZone}\label{sec:appendix_trustzone}

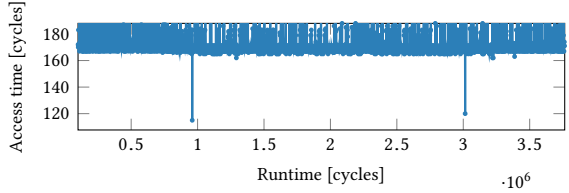
\begin{figure}[!htb]
 \centering
 \begin{tikzpicture}
\begin{axis}[
mlineplot,
style={font=\footnotesize},
xlabel={Runtime [cycles]},
ylabel={Access time [cycles]},
width=0.95\hsize,
scaled y ticks=false,
xmin=100000,
xmax=3763666,
ymax=188,
height=3cm
]
\addplot+[blue,thick,mark options={draw=blue,fill=blue},mark=*,mark size=0.5] table[x=Time,y=Access,col sep=space] {data/out_trustzone.csv};

\end{axis}
\end{tikzpicture}
 \caption{A double fetch of a trustlet running inside the ARM TrustZone of a Raspberry Pi 3. The cache hits can be clearly seen at around $0.96 \cdot 10^6$ and $3.01 \cdot 10^6$ cycles as the access time drops from \SIx{>160} cycles to \SIx{<120} cycles.}
 \label{fig:double-fetch-trustzone}
\end{figure}

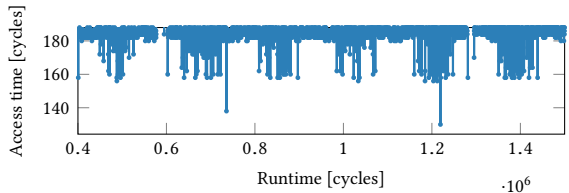
\begin{figure}[!htb]
 \centering
 \begin{tikzpicture}
\begin{axis}[
mlineplot,
style={font=\footnotesize},
xlabel={Runtime [cycles]},
ylabel={Access time [cycles]},
width=0.95\hsize,
scaled y ticks=false,
xmin=400000,
xmax=1500000,
ymax=188,
height=3cm
]
\addplot+[blue,thick,mark options={draw=blue,fill=blue},mark=*,mark size=0.5] table[x=Time,y=Access,col sep=space] {data/out_sgx.csv};

\end{axis}
\end{tikzpicture}
 \caption{Monitoring a double fetch inside an SGX enclave. The cache hits can be clearly seen at around $0.75 \cdot 10^6$ and $1.21 \cdot 10^6$ cycles as the access time drops from \SIx{>150} cycles to \SIx{<140} cycles.}
 \label{fig:double-fetch-sgx}
\end{figure}

ARM TrustZone is a trusted execution environment for the ARM platform. 
The processor can either run in the normal world or the trusted world. 
As with Intel SGX, the worlds are isolated from each other using hardware mechanisms. 
Trustlets---applications running inside the secure world---provide a well-defined interface to normal world applications. 
This interface is accessed through a secure monitor call, similar to a syscall.

To use the ARM TrustZone, the normal-world operating system requires TrustZone support. 
Furthermore, a secure-world operating system has to run inside the TrustZone. 
For the evaluation, we used the TrustZone of a Raspberry Pi 3. 
We use the open-source trusted execution environment OP-TEE~\cite{OPTEE} as a secure-world operating system. 
The normal world runs a TrustZone-enabled Linux kernel. 

As with Intel SGX (\cf \Cref{sec:eval-tees}), we again implement a trustlet providing a simple interface for receiving a pointer to a memory location. 
However, there are some subtle differences compared to the SGX enclave.  
First, trustlets are not allowed to simply access normal-world memory. 
To pass data or messages from normal world to secure world and vice versa, world shared memory is used, a region of non-secure memory, mapped both in the normal as well as in the secure world. 
With the world shared memory, we fulfill all criteria of \cref{sec:properties}. 

On ARM, there are generally no unprivileged instructions to flush the cache or get a high-resolution timestamp~\cite{arm_arch_manualv7}. 
However, they can be used from the operating system. 
Thus, in contrast to the double-fetch detection in syscalls or Intel SGX, we require root privileges to detect double fetches inside the TrustZone. 
This is not a real limitation, as we use the detection only for testing, and discovering bugs.
An attacker using \FlushReload as a trigger to exploit a double-fetch bug can rely on different time sources and eviction strategies as proposed by Lipp~\etal\cite{Lipp2016}. 

\cref{fig:double-fetch-trustzone} shows a recorded cache trace of the trustlet. 
Similarly to \Cref{fig:double-fetch-sgx}, a trace from Intel SGX, the cache hits are clearly distinguishable from the cache misses. 
Thus, we can detect double-fetch bugs in trustlets, even without having access to the corresponding binaries. 

\FloatBarrier

\section{Example of DropIt}\label{sec:appendix_dropit}

In this section, we show a small example of how to use \DropIt. 
Listing~\ref{lst:drop-it} shows an example how to protect the insecure \texttt{strcpy} function using \DropIt. 
A programmer only has to add 3 lines of code (highlighted in the listing) to protect arbitrary code from double fetch exploitation.
\DropIt is clearly simpler than current state-of-the-art software-based retry logic (\cf \Cref{alg:string-cpy}). 

\DropIt is implemented in standard C without any dependencies on other libraries, and can thus be used in user space, kernel space, as well as in trusted execution environments (\eg Intel SGX). 
If Intel TSX is not available, \DropIt has the possibility to execute a fall-back function instead.

\begin{lstlisting}[float,language=C,caption={Using \DropIt to protect a simple string copy containing a double-fetch bug from being exploited. Only the highlighted lines (1, 2, and 10) have to be added to the existing code to eliminate the double-fetch bug.},label={lst:drop-it}]
(*@\hl{\texttt{dropit\_t config = dropit\_init(1000);}}@*)
(*@\hl{\texttt{dropit\_start(config);}}@*)
len = strlen(str); // First access
if (len < sizeof(buffer)) {
  strcpy(buffer, str); // 2nd access, 
    // length of 'str' could have changed
} else {
  printf("Too long!\n");
}
(*@\hl{\texttt{dropit\_end(config, \{ printf("Fail!"); exit(-1);\});}} @*)
\end{lstlisting}

\end{document}